\documentclass[fonts]{icst}

\usepackage{moreverb}

\usepackage[breaklinks,colorlinks,bookmarksopen,bookmarksnumbered,linkcolor=ICSTblue,citecolor=blue,urlcolor=ICSTblue]{hyperref}
\usepackage{breakurl}
\usepackage{doi}

\usepackage{graphicx,flushend}

\usepackage{acronym}
\usepackage{todonotes}
\usepackage{multirow}
\usepackage{tabularx}
\usepackage[none]{hyphenat}

\graphicspath{ {./images/} }
\newcommand{\tabitem}{~~\llap{\textbullet}~~}

\newacroplural{ctef}[CTEFs]{Cybersecurity Test and Evaluation Facilities}
\newacroplural{rat}[RATs]{Radio Access Technologies}
\newacro{3gpp}[3GPP]{3rd Generation Partnership Project}
\newacro{adas}[ADAS]{Advanced Driver-assistance System}
\newacro{arcit}[ARC-IT]{Architecture Reference for Cooperative and Intelligent Transportation}
\newacro{c2c}[C2C]{Car-2-Car}
\newacro{cav}[CAV]{Connected Autonomous Vehicle}
\newacro{ca}[CA]{Certification Authority}
\newacro{cia}[CIA]{Confidentiality, Integrity and Availability}
\newacro{cits}[C-ITS]{Cooperative Intelligent Transportation System}
\newacro{cpps}[CPPS]{Cyber-Physical Production System}
\newacro{csce}[CSCE]{Cybersecurity Centre of Excellence}
\newacro{ctef}[CTEF]{Cybersecurity Test and Evaluation Facility}
\newacro{cv2x}[C-V2X]{Cellular Vehicle-to-Everything}
\newacro{dos}[DoS]{Denial-of-Service}
\newacro{faas}[FaaS]{Function-as-a-Service}
\newacro{gdpr}[GDPR]{General Data Protection Regulation}
\newacro{its}[ITS]{Intelligent Transportation System}
\newacro{maas}[MaaS]{Mobility-as-a-Service}
\newacro{mitm}[MitM]{Man-in-the-Middle}
\newacro{ml}[ML]{Machine Learning}
\newacro{ncsc}[NCSC]{National cybersecurity Centre }
\newacro{oran}[O-RAN]{Open Radio Access Network}
\newacro{oss}[OSS]{Open-Source Software}
\newacro{ota}[OTA]{Over-the-Air}
\newacro{paas}[PaaS]{Platform-as-a-Service}
\newacro{pki}[PKI]{Public Key Infrastructure}
\newacro{qos}[QoS]{Quality-of-Service}
\newacro{rat}[RAT]{Radio Access Technology}
\newacro{saas}[SaaS]{Software-as-a-Service}
\newacro{samm}[SAMM]{Software Assurance Maturity Model}
\newacro{sca}[SCA]{Software Composition Analysis}
\newacro{sdlc}[SDLC]{System Development Life Cycle}
\newacro{secaas}[SecaaS]{Security-as-a-Service}
\newacro{siem}[SIEM]{Security Information and Event Management}
\newacro{soar}[SOAR]{Security Orchestration, Automation and Response}
\newacro{sos}[SoS]{System-of-Systems}
\newacro{v2g}[V2G]{Vehicle-to-Grid}
\newacro{v2i}[V2I]{Vehicle-to-Infrastructure}
\newacro{v2p}[V2P]{Vehicle-to-Pedestrian}
\newacro{v2v}[V2V]{Vehicle-to-Vehicle}
\newacro{v2x}[V2X]{Vehicle-to-Everything}
\newacro{vanet}[VANET]{Vehicular Ad-hoc Network}

\newcommand\BibTeX{{\rmfamily B\kern-.05em \textsc{i\kern-.025em b}\kern-.08emT\kern-.1667em\lower.7ex\hbox{E}\kern-.125emX}}

\def\volumeyear{2014}

\journalname{XXXXXX}
\articletype{Research Article/Editorial}
 \def\volumeyear{XXXX}
 \def\volumenumber{XX}
  \def\issuemonth{XX-XX}
  \def\issuenumber{X}
  \def\articleid{XX}
  \def\copyrightholder{Author Name\textit{ et al.}, licensed to ICST}
  \copyrightnote{This is an open access article distributed under the terms of the Creative Commons Attribution license (\url{http://creativecommons.org/licenses/by/3.0/}), which permits unlimited use, distribution and reproduction in any medium so long as the original work is properly cited.}
  \def\articledoi{10.4108/XX.X.X.XX}
  \received{XXXX}
  \accepted{XXXX}
  \published{XXXX}

\begin{document}

\runningheads{}{}



\title{Cybersecurity in Motion: A Survey of Challenges and Requirements for Future Test Facilities of CAVs}

\author{Ioannis Mavromatis\fnoteref{1}\affilnum{1}, Theodoros Spyridopoulos\affilnum{3}, Pietro Carnelli\fnoteref{1}\affilnum{2}, Woon Hau Chin\affilnum{2}, Ahmed Khalil\affilnum{4}, Jennifer Chakravarty\affilnum{4}, Lucia Cipolina Kun\affilnum{4}, Robert J. Piechocki\affilnum{4}, Colin Robbins\affilnum{5}, Daniel Cunnington\affilnum{5}, Leigh Chase\affilnum{6}, Lamogha Chiazor\affilnum{6}, Chris Preston\affilnum{7}, Rahul\affilnum{7}and Aftab Khan\fnoteref{1}\affilnum{2}}

\address{\affilnum{1}Digital Catapult, London, UK\\
\affilnum{2}Bristol Research \& Innovation Laboratory, Toshiba Europe Ltd., Bristol, UK\\
\affilnum{3}School of Computer Science \& Informatics, Cardiff University, Cardiff, UK\\
\affilnum{4}Department of Electrical and Electronic Engineering, University of Bristol, Bristol, UK\\
\affilnum{5}Nexor Ltd., Nottingham, UK,\\
\affilnum{6}IBM Research, Winchester, UK,\\
\affilnum{7}Honda R\&D Europe (UK) Ltd., Reading, UK 
}

\abstract{
The way we travel is changing rapidly, and Cooperative Intelligent Transportation Systems (C-ITSs) are at the forefront of this evolution.  However, the adoption of C-ITSs introduces new risks and challenges, making cybersecurity a top priority for ensuring safety and reliability. Building on this premise, this paper presents an envisaged Cybersecurity Centre of Excellence (CSCE) designed to bolster research, testing, and evaluation of the cybersecurity of C-ITSs. We explore the design, functionality, and challenges of CSCE's testing facilities, outlining the technological, security, and societal requirements. Through a thorough survey and analysis, we assess the effectiveness of these systems in detecting and mitigating potential threats, highlighting their flexibility to adapt to future C-ITSs. Finally, we identify current unresolved challenges in various C-ITS domains, with the aim of motivating further research into the cybersecurity of C-ITSs.
}



\fnotetext[1]{Corresponding authors: Ioannis.Mavromatis@digicatapult.org.uk, \{Pietro.Carnelli, Aftab.Khan\}@toshiba-bril.com}

\maketitle

\section{Introduction}
\acp{cits} and the \acp{cav}, are expected to revolutionise the \ac{maas} paradigm~\cite{itsRevolution}. As we move towards smarter and more sustainable cities, there will be numerous opportunities for \ac{maas} use-cases in areas such as on-demand transportation, accessibility, and road safety~\cite{useCases}. However, these use cases will require specialised hardware, complex software implementations, and scalable data architectures~\cite{mainPlanes}. \acp{cits}  integrate multiple entities for enhanced transport solutions. As they grow, \acp{cits} become complex \ac{sos}, where independent systems combine to achieve broader functionalities. This complexity introduces significant cybersecurity risks. The different ``data surfaces'', communication interfaces, and Internet-facing entry points for such an ecosystem increase the potential vulnerabilities and attack surfaces within a \ac{cits}~\cite{attackSurfaces}. This is especially concerning as the increase in vehicle autonomy means a single attack point could have catastrophic effects on the entire \acp{cits} ecosystem and the fleet of vehicles~\cite{singlePoint}.

As CAV adoption increases, a robust cybersecurity assurance framework is needed to ensure safety and public trust~\cite{assuranceCAVs}. Therefore, this paper presents the concept of a \ac{ctef} targeting \acp{cits} and a \ac{cav}-enabled \ac{csce} and a survey of the existing challenges and requirements for such frameworks. \ac{ctef} is meant to provide comprehensive testing, certification and monitoring services for \acp{cav}. The cybersecurity considerations of a \ac{cits} should be tackled holistically~\cite{cybersecurityHolisticallyTackled}. Our proposed solution revolves around cybersecurity testing, evaluation and certification capabilities for future \acp{cav} and related \ac{cits} infrastructure (e.g., future roadside \ac{cav} traffic coordination units), as well as provisions for live attack monitoring of future \ac{cits} implementations. 

The proposed framework includes components for in-vitro, i.e., separate/``glass-walled'', testing facilities in isolated environments, in-situ, i.e., ``on-the-premises'', realistic hardware/software evaluation using simulated environments, and in-vivo, i.e., real-world or live testing conditions in actual operational scenarios. It should also support scalable and extensible architectures, cybersecurity assessment schemes, and real-time monitoring of threats. \ac{ctef} is intended to investigate cybersecurity vulnerabilities and threats within a sub-system, system, or \ac{sos} fashion. To achieve the above, both virtual and physical test facilities are required. 

\begin{table}[ht] 
\renewcommand{\arraystretch}{1.05}
\centering
    \caption{List of Acronyms.}
    \begin{tabular}{lp{0.68\columnwidth}}
        \toprule
        \textbf{Acronym} & \textbf{Description} \\
        \midrule
        3GPP & 3rd Generation Partnership Project \\
        ADAS & Advanced Driver-assistance System \\
        ARC-IT & Architecture Reference for Cooperative and Intelligent Transportation \\
        C2C & Car-2-Car \\
        CAV & Connected Autonomous Vehicle \\
        CA & Certification Authority \\
        CIA & Confidentiality, Integrity and Availability \\
        C-ITS & Cooperative Intelligent Transportation System \\
        CPPS & Cyber-Physical Production System \\
        CSCE & Cybersecurity Centre of Excellence \\
        CTEF & Cybersecurity Test and Evaluation Facility \\
        C-V2X & Cellular Vehicle-to-Everything \\
        DoS & Denial-of-Service \\
        FaaS & Function-as-a-Service \\
        GDPR & General Data Protection Regulation \\
        ITS & Intelligent Transportation System \\
        MaaS & Mobility-as-a-Service \\
        MitM & Man-in-the-Middle \\
        ML & Machine Learning \\
        NCSC & National Cybersecurity Centre \\
        O-RAN & Open Radio Access Network \\
        OSS & Open-Source Software \\
        OTA & Over-the-Air \\
        PaaS & Platform-as-a-Service \\
        PKI & Public Key Infrastructure \\
        QoS & Quality-of-Service \\
        RAT & Radio Access Technology \\
        SaaS & Software-as-a-Service \\
        SAMM & Software Assurance Maturity Model \\
        SCA & Software Composition Analysis \\
        SDLC & System Development Life Cycle \\
        SecaaS & Security-as-a-Service \\
        SIEM & Security Information and Event Management \\
        SOAR & Security Orchestration, Automation and Response \\
        SoS & System-of-Systems \\
        V2G & Vehicle-to-Grid \\
        V2I & Vehicle-to-Infrastructure \\
        V2P & Vehicle-to-Pedestrian \\
        V2V & Vehicle-to-Vehicle \\
        V2X & Vehicle-to-Everything \\
        VANET & Vehicular Ad-hoc Network \\
        \bottomrule
    \end{tabular}\label{tab:acronyms}
\end{table}

Overall, this paper makes recommendations in four key areas - testing facilities, architectures, assessment schemes and ecosystem requirements. Based on an extensive study of existing best practices and requirements, we describe the methodology for achieving the \ac{ctef}'s vision and establishing a rigorous cybersecurity assurance program for \acp{cav}. Our work discusses good practices, standards, considerations, and technological aspects (architectures, technologies and techniques) to accelerate the safe and secure development, trialling, testing and deployment of a cybersecure \ac{cits}. 

Like any other \acp{csce} found in smart cities or smart factories, key principles like \textit{futureproofing}, \textit{scalability}, \textit{extensibility}, \textit{modularity}, \textit{flexibility} and \textit{reliability} need to be considered for such a framework~\cite{keyPrinciples}. The integration of new tools will enable continuous enhancement of the \ac{ctef}'s capabilities. Architectures based on concepts like fog computing~\cite{agile-data-offloading-novel-fog-computing}, virtualisation~\cite{virtualisation} and cloud-native/serverless computing~\cite{Kem15} are critical. Our work discusses their importance, adoption and integration within a unified framework. Detailed test regimes combining automated and manual testing are also described. Assurance frameworks aligned with standards like ISO 21434~\cite{ISO21434} are discussed. Finally, the ecosystem needs of the \ac{ctef} in terms of expertise, infrastructure, and collaboration with regulators and academia are also highlighted.

This paper is structured as follows. Sec.~\ref{sec:background} explores the foundational aspects of \acp{cav}, describing the data and attack surfaces and the main system components found in \acp{cits}. Related articles are described in Sec.~\ref{sec:relatedWork}, providing insights into the existing literature. The core idea behind the paper is presented in Sec.~\ref{sec:csceSystem}, where we detail the requirements and its scope. Sec.~\ref{sec:systemDesign} further discusses the architectural design of the \ac{csce}. The test facilities' design and requirements are described in Sec.~\ref{sec:designCTEF}, explaining the in-vitro and in-situ testing and analysis. Sec.~\ref{sec:technologies} presents the core technologies required within \ac{csce} and aligns them with the requirements introduced earlier. The steps that should be taken for smooth integration with the real world are commented on in Sec.~\ref{sec:inVivo}. The challenges identified across the various entities within \ac{csce} are discussed in Sec.~\ref{sec:challenges}, addressing potential hurdles and considerations for a real-world integration and emphasising aspects like privacy and operational requirements. Sec.~\ref{sec:conclusion} concludes the paper and summarises our key findings and contributions. Finally, a table summarising all the acronyms can be found in Tab.~\ref{tab:acronyms}.

\section{Background}\label{sec:background}
\acp{cits} refers to transport systems where the cooperation between two or more sub-systems (e.g., personal, vehicle, roadside and central) enables and provides enhanced services, compared to the traditional \acp{its}. \acp{cits} utilise wireless communication links to enable real-time \ac{v2x} connectivity. This, in turn, enables far greater coordination between road users and the involved systems and creates safer and more efficient traffic flows~\cite{shuttlesCyber}. 

A \ac{cits} reference architecture provides a common framework for planning, defining, and integrating all system components~\cite{referenceArchitectureToUse}.  It essentially provides a common basis for planners and engineers with differing concerns to conceive, design and implement systems using a ``common language''.  Most critically, reference architectures can help perform attack surface analysis, identify threats, and understand \textit{how} an attack could be executed. 

\begin{figure*}[t]     
\centering
\includegraphics[width=0.7\textwidth]{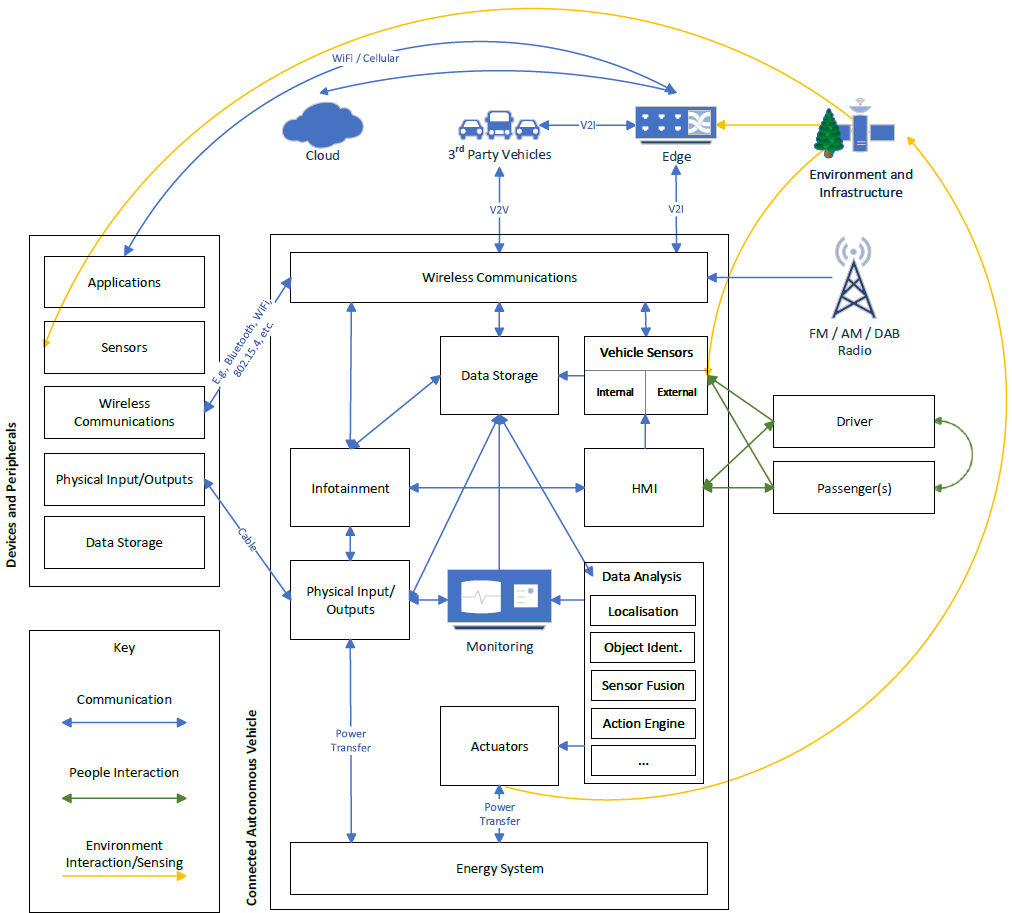}
    \caption{\ac{cav} and devices \& peripherals reference architecture~\cite{referenceArchitectureToUse}.}
    \label{fig:referenceArchitecture}
\end{figure*}

The literature provides various reference architectures.  Some lack important details to derive certain attack categories~\cite{referenceArchitecture1}, while others are too intricate to interpret by vehicle manufacturers and CAV system designers~\cite{referenceArchitecture2}.  For our work, we use the reference architecture from~\cite{referenceArchitectureToUse} as a basis, also shown in Fig.~\ref{fig:referenceArchitecture}.  This reference architecture depicts the main components for \acp{cav} and the devices and peripherals that interact with a \ac{cav}, proposing a hybrid Functional-Communication viewpoint that balances the interactions' complexity and depth.  More details about the architecture and a description of the components can be found in~\cite{referenceArchitectureToUse}.

\subsection{CAV Attack Surface}\label{subsec:attackSurface}
To identify attack surfaces that a threat agent would exploit, two questions should be considered, i.e., what a \ac{cav} does and how the CAV can be interacted with to be attacked.  As depicted in Fig.~\ref{fig:referenceArchitecture}, various communication buses and links across multiple physical and virtual devices and entities span both the ``Cloud'' and ``Edge'' layers.

Following that, we can define an attack surface that constitutes multiple entry points or attack vectors that a malicious user could exploit to gain control, enter or extract data, or perform other malicious activity within a \ac{cits}~\cite{cavAttackSurfaces}.  Any device, system, software, or actor (as shown in Fig.~\ref{fig:referenceArchitecture}), both internal and external, that communicates with a \ac{cav} component contributes to the attack surface. 

The ``\textbf{digital data surface}'' underpins the attack surface.  This constitutes the individual data points that flow between system components.  The system components will transfer data across the ``surface'' that could be vulnerable to manipulation and attack.  We broadly refer to the ``data surface'' as the data points generated by CAV platforms and flow between local and networked components~\cite{mainPlanes} (\ac{v2x}).  Finally, a single ``data flow'' could be described as the flow of information between two specific endpoints (blue lines in Fig.~\ref{fig:referenceArchitecture}).

\subsection{Modelling and Representing the Data Surface}\label{subsec:modellingData}
The ``data surface'' resides ``\textit{above}'' in the CAV platform, providing a logical representation of the platform. In other words, it could be described as a ``data representation'' of the system platform as seen from the network perspective. This abstraction is crucial for understanding the flow of data, its sources, and its consumers within the CAV ecosystem.

\ac{cav} systems can be considered as individual entities or \acp{sos} -- the latter being a way to group entities that provide a common service or function~\cite{dataModelling}. An individual \ac{cav} platform can be described in terms of:
\begin{itemize}[left=0em]
    \item \textbf{Systems}: a collection of interconnected elements (hardware, firmware, software, etc.) that, combined, achieve specific functionality (or well-bounded set of functions). For instance, a navigation system in a \ac{cav} might combine GPS hardware, map software, and route optimisation algorithms. 
    \item \textbf{Sub-systems}: individual instances of hardware, firmware and software that can be aggregated to form a system. For example, the GPS module can be a sub-system within the above-mentioned navigation system.
    \item \textbf{Devices}: components within sub-systems that generate or consume the actual data flowing through the CAV platform. The data could be simple data types such as numbers, strings, or more complex structures.
\end{itemize}

A system is a collection of sub-systems that provide a specific \textit{service} or \textit{capability}. The data systems can be described broadly as: 1) \textit{producers}: entities that yield data, such as sensors or the output of processing functions; 2) \textit{consumers}: entities that take data as input, such as applications components or the input to processing functions.

The data surface can be expressed geometrically across $d$ dimensions with $n$ data points. It is important to take into account the data types and structures (e.g., simple numeric variables or complex XML structures) as we create layers in the model for different types of data interactions. For instance, raw sensor data might be on one layer, processed data on another, and application-level data interactions on a third. Finally, for a more detailed model, one should include information about the time and frequency domains of the data interactions. This can help in understanding real-time requirements and optimising data flow for efficiency. Some more information about the above can be found in~\cite{dataModelling}, described from the point of view of a Smart City scenario.

\subsection{C-ITS Communication Domains}\label{subsec:communicationDomains}
Having defined an attack surface and how the data are represented and modelled, we can later identify potential attack vectors, considering the different communication domains.  The \ac{c2c} Communication Consortium identifies three key domains~\cite{surveyCAVS}:
\begin{itemize}[left=0em]
    \item \textit{Intra-vehicle domain}: The different systems and components inside the vehicle (e.g., sensors, onboard computers, infotainment system, \ac{adas} features, powertrain, etc.).
    \item \textit{Ad-hoc domain}: The \ac{v2v} and \ac{v2i} communication systems that allow real-time data exchange between vehicles and other entities on the road.
    \item \textit{Infrastructure domain}: The fixed infrastructure systems like traffic management centres, roadside sensor networks, edge computing nodes, cellular base stations, etc., that provide connectivity support, traffic management, emergency response and other \ac{cits} services.
\end{itemize}

As \acp{cits} are getting more complex, the above has been further extended recently, adding:
\begin{itemize}[left=0em]
    \item \textit{Central cloud domain}: The remote cloud computing infrastructure that provides additional services and capabilities to vehicles and transportation infrastructure.  For example, it can offer intelligent decision-making agents for traffic management~\cite{motionPlanning}.
    \item \textit{Personal domain}: The mobile devices, wearables and other gadgets carried by individuals that can connect to vehicles and infrastructure in a \ac{v2p} fashion~\cite{v2p}.
    \item \textit{Enterprise domain}: The third-party service providers and businesses that are stakeholders in the \ac{cits} ecosystem~\cite{ISO17427-1}.  For example, logistics companies managing fleets, insurance providers, location-based service companies, etc.
\end{itemize}

Sophisticated attacks may target multiple \ac{cav} components across various domains.  Therefore, a single attack could exhibit multiple data flows across the data surface.  In terms of potential \ac{cav} cyber-attacks that constitute the attack surface, an attacker may exploit vulnerabilities within the physical infrastructure (e.g., electric charging stations), perform adversarial modifications to sensor data~\cite{le3d,le3dDemo}, perform poisoning attacks to intelligent agents~\cite{conceptDrift}, or perform network attacks that disrupt communications (e.g., a \ac{dos} attack~\cite{citsAttacks}).  For a comprehensive overview of various attack points that constitute the \ac{cav} attack surface, we refer the reader to~\cite{referenceArchitectureToUse,shuttlesCyber,attacks1,attacks2}.

\subsection{Security Objectives for a C-ITS}\label{sub:securityObjectives}
From a cybersecurity standpoint, and based on the above, the five main goals to be achieved are:
\begin{enumerate}[left=0em]
    \item To model and understand potential attack vectors in complex \ac{cav} and \ac{cits} architectures.
    \item To detect malicious network interactions, differentiating these amongst a high volume of valid interactions.
    \item To distinguish tampered data from normal data and identify attacks in intelligent agents.
    \item To protect \acp{cav} and \acp{cits} through the deployment of detection and mitigation techniques, limiting the impact of a cyber-attack given the interconnected nature of \ac{cav} systems.
    \item To ensure the low-latency operation of the deployed detection and mitigation functionality.
\end{enumerate}

The above points have been demonstrated in practice in the past. For example, authors in~\cite{jeepCherokee} demonstrated how they exploited a vulnerability in the vehicle's infotainment system (connected via a cellular network) to remotely take control of the vehicle (goal no.1). In~\cite{falseDataInjection}, a false data injection attack is demonstrated and the authors provide a way to detect and isolate the attack from regular data exchanged within a \ac{cav} application (goal no.2). A sensor fusion technique and a model to extrapolate the vehicle's position and project projection were used in~\cite{sensorSpoofing} to mitigate against tampered speedometer data (goal no.3). The importance of mitigation strategies is discussed in~\cite{overTheAirUpdate}, describing how \ac{ota} updates can help manufacturers remotely patch vulnerabilities and reduce the potential impact of cyber-attacks (goal no.4). Finally, in~\cite{lowLatencyDrive}, the importance of reliability and reduced overhead is discussed proposing a low-overhead model to identify malicious roadside basestations (goal no.5).

The exchange of data and the various wireless interfaces increase the potential attack vectors within a \ac{cits}. These threats can expose critical traffic systems and compromise the safety of all passengers and pedestrians. Therefore, it is essential to introduce a solid security framework. \ac{arcit}~\cite{arcit} describes this framework in the context of three cybersecurity pillars, i.e., \ac{cia}. Briefly, the \ac{cia} defines:
\begin{enumerate}[wide, labelwidth=!, labelindent=9pt]
    \item \textbf{\textit{Confidentiality}}: Restricts access to sensitive information based on the type of information disclosed. For example, confidential data from individual \acp{cav} should be protected as it can be misused. Vehicle ID and speed data can easily identify if the vehicle violates speed limits~\cite{confidentiality}. Therefore, \acp{cav} should use infrastructure network services to preserve anonymity.
    \item \textbf{\textit{Integrity}}: Guarantees the reliability and accuracy of the information and messages exchanged while preventing the alteration of the data from unauthorised intentions or authorised but unintended acts. For example, a sensor fusion mechanism could be used to avoid data modification. Abnormalities could be detected based on the information collected from different but complementary sensors (e.g., camera and LiDAR data). Later, the data could be either discarded or sanitised before being used~\cite{integrity}.
    \item \textbf{\textit{Availability}}: Ensures authorised access to critical information and system. Availability aims to allow access rather than restrict it. For example, traffic information should be publicly available to all \acp{cav} at any given time~\cite{availability}. 
\end{enumerate}

These three are the most crucial components of cybersecurity and form a model to guide policies for \ac{cits} information security. Balancing between them, we can ensure high-quality security standards and policies without compromising the usability of a \ac{cits}. All cybersecurity services support one or more of these objectives. Similarly, all threats undermine one or more of these objectives. Comparing security systems using these metrics can aid designers in selecting between alternatives. 

Trust and privacy are two vital aspects of \acp{cits} cybersecurity. Trust is achieved by ensuring the exchanged information's integrity, authenticity, and confidentiality. On the other hand, privacy requires rigorous measures to protect personal data from unauthorised access and potential misuse. ETSI TS-102941~\cite{etsi10941} highlights the significance of these principles in information exchange within transportation systems. The technologies integrated into \acp{cav} and \acp{cits} provide a way to track individuals and vehicles with heightened precision.  Maintaining data privacy, such as the license plate or vehicle owner, becomes essential. For example, \ac{gdpr}\footnote{GDPR official website: https://gdpr-info.eu/} establishes stringent data protection and privacy guidelines for individuals and serves as a gold standard for data handling worldwide. Adhering to \ac{gdpr} provisions, a \ac{cits} can reinforce user trust, ensure compliance, and ensure the fundamental right of privacy for travellers, fortifying their confidence in the system.

\subsection{C-ITS Concepts and System Design}\label{sub:citsSystemDesign}
\begin{figure}[t]     
\centering
\includegraphics[width=1\columnwidth]{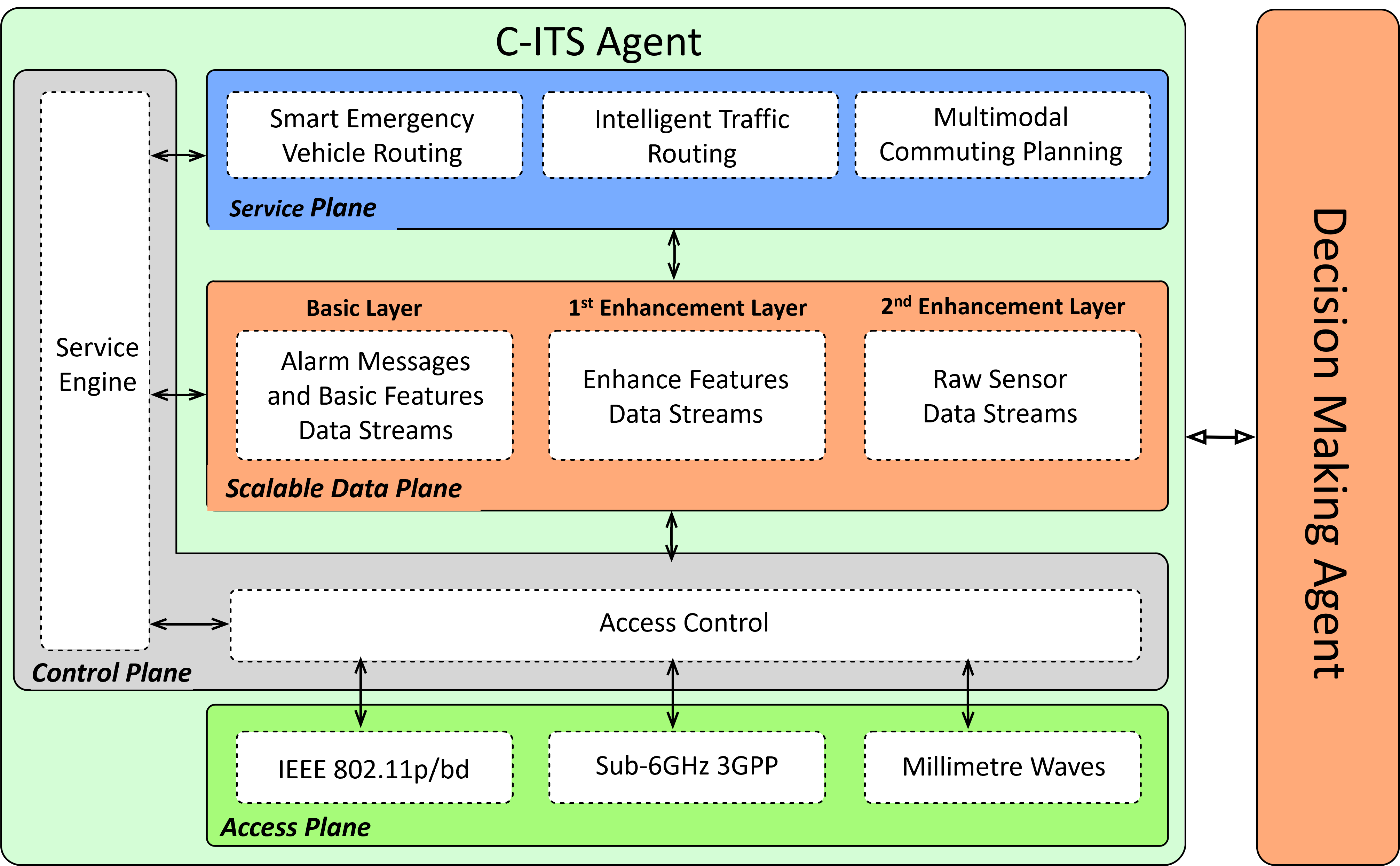}
    \caption{The high-level system design of an \ac{cits} system, showing the interaction between the communication domains, the data exchanged and the different services.}
    \label{fig:itsAgent}
\end{figure}

Sec.~\ref{subsec:communicationDomains} discussed the communication domains within a \ac{cits}. A \ac{cits}, being an \ac{sos}, is responsible for: i) handling scalable applications consisting of independent data flows, ii) employing multiple \acp{rat} for each flow and mapping them on different layers according to target \ac{qos} constraints, iii) using intelligent agents for decision making. Fig.~\ref{fig:itsAgent} shows an example of such a system, demonstrating the different data, control and access planes and some example services.

As seen in the literature, future \ac{cits} are meant to provide different types of services, each one with its own \ac{qos} requirements~\cite{citsServicesClassification,mainPlanes}. This could be achieved by providing enhancement layers, extending the base layer's functionality and fulfilling the scalability and extensibility requirements. For example,~\cite{mainPlanes} proposes three main communication planes: i) a base layer, usually based on IEEE 802.11p/bd, responsible for base safety critical message exchange, ii) an initial enhancement layer, based on sub-6GHz \ac{3gpp} standards and \ac{cv2x} technologies, that can support services spanning over large geographical areas, iii) a second enhancement layer, based on Millimetre-Waves (using IEEE 802.11ad/ay or mmWave 5G NR), that fulfils the requirements for very high data rate and very low latency of the future \ac{cits} services (also seen in Fig.~\ref{fig:itsAgent}). These technologies and the different layers align with the principle of a scalable data infrastructure mentioned before. However, as described in Sec.~\ref{subsec:communicationDomains}, its one becomes an attack vector that should be taken into account within a \ac{cits}.

\begin{figure*}[t]     
\centering
\includegraphics[width=1\textwidth]{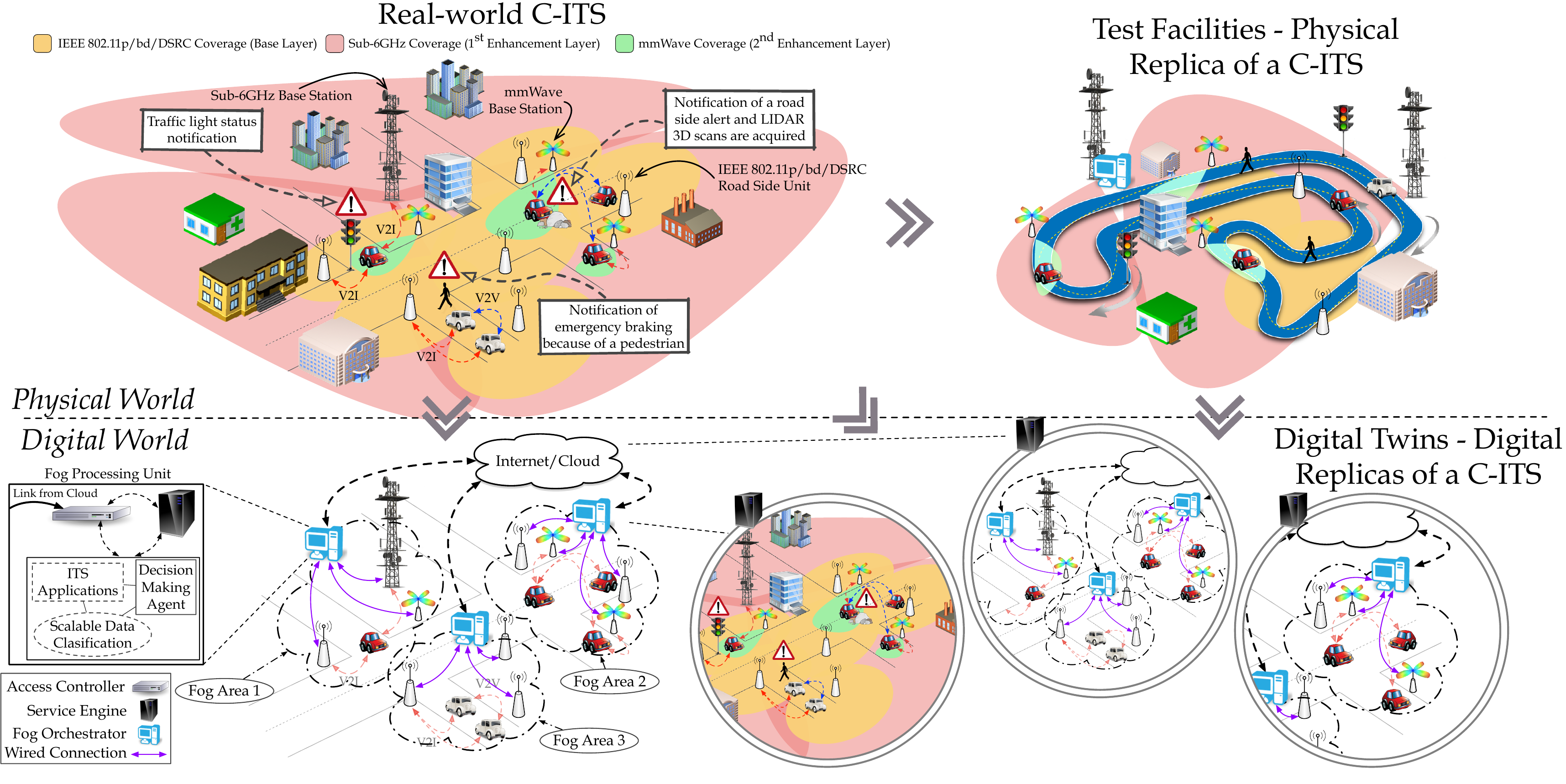}
    \caption{General overview of the considered \ac{cits} model and the connection with \ac{csce}. Physical and Digital \ac{cits} Replicas will be provided within \ac{csce} to enhance the cybersecurity and \ac{cits} operation.}
    \label{fig:citsSystem}
\end{figure*}

The computing infrastructure supporting a \ac{cits} is of paramount importance. It supports large-scale \ac{cits}-related applications such as road safety, traffic efficiency, multimodal commuting, smart parking, etc. For time-critical applications, a hybrid computing model using Cloud, Edge, and Fog paradigms~\cite{mainPlanes,citsServicesClassification} for data analytics and knowledge discovery is required (left-hand side of Fig.~\ref{fig:citsSystem}). Bringing the computing resources closer to the ``Edge'' enables faster processing and data collection and minimises the network delays introduced by the several hops in the backbone connectivity. Edge processing capabilities are particularly important for \ac{ml}-based solutions, as different types and quantities of machines can work together to accomplish specific objectives. Secs~\ref{subsec:attackSurface} and~\ref{sub:securityObjectives} briefly touched upon the attack surfaces related to the data plane and the importance of concrete security solutions.

Collected data should be managed appropriately in a \ac{cits}~\cite{Kem15}. Data analysis or mining at the Edge can be performed in different ways. Usually, long-term data storage is handled by the Cloud plane. On the other hand, data that help with time-critical applications are stored in Edge/Fog nodes~\cite{agile-data-offloading-novel-fog-computing} that are designed to be always ready to use (Fig.~\ref{fig:citsSystem}). The acquired \ac{cits} data are usually processed by first inspecting their correctness by reviewing the type of data, error rectification, and data cleansing. The evaluated data is then analysed by sophisticated algorithms, either rule-based or ML-based. Inconsistencies during processing are fixed, and amended data can be further analysed to derive information from it. \ac{cits} data transmission and analysis at the Fog offers several advantages, these being:

\begin{itemize}[left=0em]
    \item \textbf{\textit{Low-latency services:}} Analysing data at the collection source reduces the latency as the time needed to transmit is minimised.
    \item \textbf{\textit{Resource Management:}} Vehicles or pedestrians (nodes) can join and leave the Fog plane at any time. Therefore, a high-speed resource management service will enable real-time network monitoring and control. 
    \item \textbf{\textit{Bandwidth Management:}} Due to the reduced data transmission, the available bandwidth can be better utilised for other purposes.
    \item \textit{\textbf{Location Awareness:}} Location services help monitor and track the Fog nodes. This will enable the distributed fog nodes to form a multicast group to facilitate rapid decisions for \ac{cits} application.
\end{itemize}

The above could be pivotal in achieving a fully functional \ac{cits}. However, as the above sections show, a \ac{cits} is a very complex \ac{sos} with many moving parts and attack vectors. Moving from theory to practice and bringing \acp{cits} and \acp{cav} closer to the real world requires rigorous cybersecurity testing and well-established frameworks to ensure the solutions' validity. Based on that, in the following sections, we delve into the importance and requirements for cyber test facilities -- their architecture and the supporting ecosystem -- for enabling the protection of large vehicle fleets and the future \acp{cits}.

\section{Related works}\label{sec:relatedWork}
Research activities and industrial initiatives show significant advancements in the context of \acp{cav} and \ac{cits}. Authors in~\cite{testincCertification} discussed the necessity for testing and certification for autonomous vehicles with a focus on cybersecurity and \ac{ml}, highlighting the importance of testing sensors, actuators, and software running on \acp{cav}. They also discuss the importance of accreditation schemes, the creation of vulnerability reporting databases, and how public-private partnerships should standardise testing regimes and processes. Real-world testbeds and facilities like UTAC Millbrook-Culham\footnote{UTAC Millbrook Testbed: \url{https://www.utac.com/}}, ASSURED CAV\footnote{ASSURED CAV: \url{https://www.horiba-mira.com/assured-cav/}} and Testregion DigiTrans\footnote{DigiTrans: \url{https://www.digitrans.expert/en/}} already provide facilities for testing novel \ac{cav} and \ac{cits} services. Our work aims to provide a framework such that facilities like the aforementioned can become more cybersecurity-aware, incorporating ways of testing and evaluating the cybersecurity of the services developed.

Similar facilities and \acp{csce} have been proposed in the past. An example is~\cite{HSL}, where a cybersecurity research facility is presented. This facility focuses more broadly on general-purpose sensitive data cybersecurity experiments. Their lessons learned discuss the importance of disaster recovery, continuous updates and improvement of the testing facilities and proper management of participants and resources. The team in~\cite{cyberSecLab} advocates the importance of nurturing the cybersecurity workforce and presents a framework to improve the skills of potential cybersecurity actors. They highlight the importance of data-driven cybersecurity and various tools and technologies (e.g., cloud-native computing, containers, monitoring and intrusion detection tools, etc.) necessary for cybersecurity practitioners. Building upon the same ideas, our work will focus on the requirements of \ac{cits}-related test facilities, the technologies required to provide cybersecurity testing functionality and how the outcomes can be adopted in the real world.

Finally, existing works, e.g.,~\cite{safetyStandardisation,standardsInVehicles,standardWireless} highlight the necessary evolution in the automotive and related standardisation landscape while providing ethics guidelines and upcoming regulations. Other works~\cite{shuttlesCyber} focus on the challenges and threats faced by \acp{cav}. We extend both concepts by discussing additional challenges originating from the real-world adoption of the different technologies and frameworks, identifying existing drawbacks in current standardisation activities. Finally, we touch upon the challenges arising from the operational requirements and the resilience required for a real-world system.

\begin{figure*}[t]     
\centering
\includegraphics[width=1\textwidth]{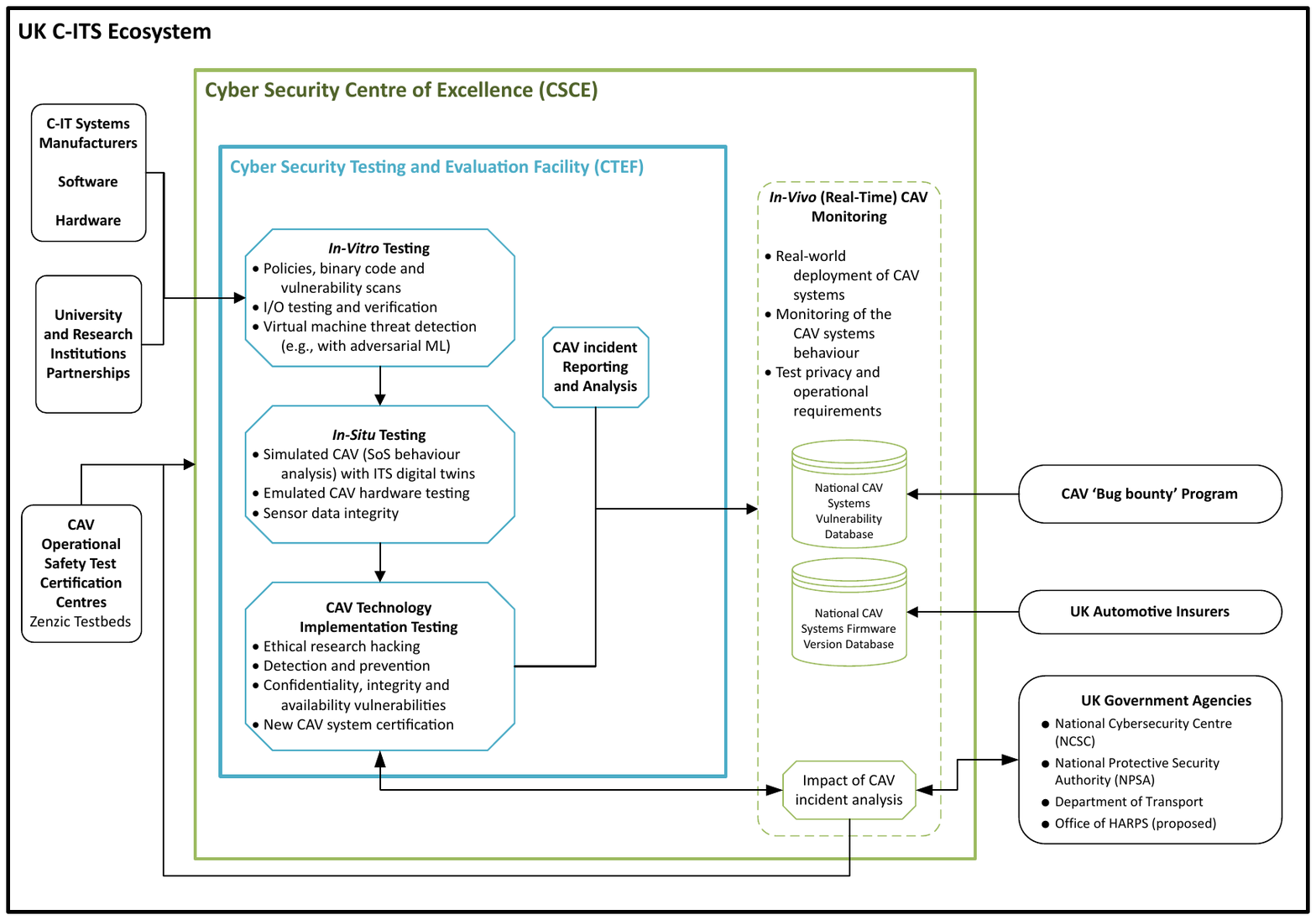}
    \caption{Proposed Cybersecurity Centre of Excellence (\ac{csce}, green box) organisational and operational structure within the UK Cooperative Intelligent Transportation System (\ac{cits}, black box). Within the \ac{csce}, we propose a cybersecurity Testing and Evaluation Facility (\ac{ctef}, blue box) to allow for online (i.e., cloud-based software containerisation of \ac{cav} systems and \ac{cits} infrastructure emulators) testing. Live monitoring of \acp{cav} operating in the UK combined with national \ac{cav} systems vulnerability and firmware version databases allows for rapid responses to security-related breaches.}
    \label{fig:csce}
\end{figure*}

\section{CSCE and CTEF System Description} \label{sec:csceSystem}
In the previous sections, we briefly described an envisaged \ac{cits} and the key security objectives for that. A \ac{csce} should build on these principles and provide the cybersecurity functionality required of a \ac{cits}. Our envisaged \ac{csce} and its corresponding sub-systems, testing regimes, and relevant stakeholders ecosystem are shown in Fig.~\ref{fig:csce}. The governmental agencies in the diagram are UK-based, but similar organisations are found in all countries and can be part of a given \ac{csce}.

\subsection{CSCE and its Testing Facilities Requirements}\label{subsec:requirements}
Our envisaged \ac{csce} should provide two main functionalities. i.e., a testing, evaluation and certification facility and a provision for live monitoring and detecting threats in real-time within a \ac{cits}. The design and proposed solutions should be replicable in the real world with minor modifications. The testing facilities should provide the tools, environments and frameworks for investigating cybersecurity vulnerabilities and threats within a sub-systemic, systemic, or \ac{sos} fashion, either in physical or digital infrastructures. This allows future \ac{cits} components and \ac{cav} manufacturers to run experiments within correctly set-up environments and test their proposed systems for cybersecurity vulnerabilities.

From the above sections, we can identify several requirements. The primary aim of the proposed \ac{csce} is to provide cybersecurity research, test and evaluation services for \acp{cits} in a \textit{scalable, adaptable, modular and flexible} way. To achieve this goal, the \ac{csce} should adhere to the following key requirements :
\begin{enumerate}[wide, labelwidth=!, labelindent=9pt, start=1,label={\bfseries CR\arabic*}]
    \item \textbf{\textit{-- Identification and validation:}} \ac{cits}-wide cyber threats must be identified and evaluated for their significance, likelihood and impact. \label{req:cscereq1}
    \item \textbf{\textit{-- Convergence of virtual and physical world:}} Both physical (e.g., conducted on a test track in an air-gapped scenario) and virtual (e.g., performed employing simulations and Digital Twins~\cite{digitalTwin1}) must be provided, with software and hardware testing support, i.e.:\label{req:cscereq2}
    \begin{itemize}[leftmargin=17pt]
        \item \textit{In-vitro testing}, i.e., ``sand-boxed'' virtual environments to test \ac{cits} systems and components.
        \item \textit{In-situ testing}, i.e., testing \ac{cits} systems interactions with each other and with {cits} related road-side infrastructure.
    \end{itemize}
    \item \textbf{\textit{-- Integration of cybersecurity tools and technologies}}: Various tools should be utilised to satisfy the safety, security and availability of \ac{cits} systems, e.g., tools for penetration testing activities or cloud-native security should be supported and provided. \label{req:cscereq3}
    \item \textbf{\textit{-- Maturing cybersecurity schemes:}} The proposed cybersecurity mechanisms and the tools developed should be matured within cybersecurity incubators. The different components must align with the \ac{cits} operational requirements, before integrated into the main system.\label{req:cscereq4}
    \item \textbf{\textit{-- Automated content curation process:}} Bidirectional communication channels should be established that feed data and information from the real world to the test facilities and continuous integration mechanism for deploying the implemented solutions.\label{req:cscereq5}
    \item \textbf{\textit{-- Continuous monitoring facilities:}} The cybersecurity incidents must be effectively detected, mitigated and prevented in real-time.\label{req:cscereq6}
    \item \textbf{\textit{-- Respond and recover mechanisms:}} Fallback policies should be in place, so when an attack occurs, to immediately begin recovery. \label{req:cscereq7}
    \item \textbf{\textit{-- Improved resilience:}} A \ac{cits} must be operational even when affected by a cybersecurity event or attack.\label{req:cscereq8}
\end{enumerate}
	
Following the requirements of the \ac{csce}, the supporting test facilities will be used as an ``evaluation and testing'' platform in an isolated fashion so that detection, mitigation and prevention mechanisms can be tested without affecting a real-world system. The following key requirements are recommended:
\begin{enumerate}[wide, labelwidth=!, labelindent=9pt, start=1,label={\bfseries TR\arabic*}]
    \item \textbf{\textit{-- Extensibility:}} All systems and subsystems must be extensible to allow the addition and testing of new security approaches, evaluation of new modules and security prototypes. Breaking down the testing into smaller manageable components makes it easier to achieve a highly secured system.\label{req:ctefreq1}
    \item \textbf{\textit{-- Distributed capabilities:}} Distributed processing capabilities must be provided to ensure scalable solutions and improve efficiency and performance.\label{req:ctefreq2}
    \item \textbf{\textit{-- Bidirectional Interaction with Intelligent Agents:}} The testing platform should handle the training data and be able to apply policies generated by \ac{ml} solutions. A set of good security policies and practices should follow, e.g., regular checks, balances, and reminders, confirming that a new security policy is being enforced.\label{req:ctefreq3}
    \item \textbf{\textit{-- Security-as-a-Service:}} The testing pipeline must employ reconnaissance and penetration testing capabilities as well as vulnerability detection mechanisms in a \ac{secaas} fashion. This provides the flexibility to test different components dynamically and introduces a softwarisation and virtualisation approach into the test facilities.\label{req:ctefreq4}
    \item \textbf{\textit{-- High Adaptability:}} The test infrastructure and pipeline must be built on a core baseline but be highly adaptable. The security level of one security domain can be adjusted without affecting the security levels of other domains or systems:\label{req:ctefreq5}
    \begin{itemize}[left=0.5em]
        \item The choice to conduct specific experiments that use individual systems, or \ac{sos}, represent specific architectures and work on different scenarios and use cases. \item Similar process activities must be grouped at the capability level. The capability level is used to assess the risk exposure of assets and processes and specify adequate and consistent security requirements.
        \item The testing platform must provide tools for threat detection of common network vulnerabilities and aggregated and systemic threats across large fleets of vehicles or \ac{cits} components.
        \item The test framework should be easily reset to the stable core functionality and be prepared for a new set of experiments. Significant security risks introduced during the experimentation of a system should be easily reverted back to the state that was proven to be more secure.
    \end{itemize}
    \item \textbf{\textit{-- External cooperation:}} The test facilities must facilitate cooperation with the industry, academia and government to maintain and improve security standards. Using defined security domains allows organisations to engage business partners in determining the appropriate security requirements for each cross-organisational information flow.\label{req:ctefreq6}
    \item \textbf{\textit{-- Existing development practices}}: The test facilities must utilise existing well-established development practices but also allow collaborations with R\&D to ensure up-to-date testing and state-of-the-art equipment and technologies are always used.\label{req:ctefreq7}
\end{enumerate}

\subsection{Scope and Key Characteristics of CSCE}
The envisaged \ac{csce} should be able to address the pressing concerns around the cyber threats for \acp{cits} and \acp{cav}. It should address both the technological risks as well as provide suggestions for the socio-technical aspects of the surrounding ecosystem (e.g., skills required, public engagement, training on secure practices, etc.). Three different entities will support the \ac{csce} (Fig.~\ref{fig:csce}):

\begin{enumerate}[wide, labelwidth=!, labelindent=9pt]
    \item \textbf{\textit{Cybersecurity Test and Evaluation Facility (CTEF)}}: \ac{ctef} is a large and complex \ac{sos} based on legacy and state-of-the-art technologies. Like all industrial control systems, \ac{ctef} will have unique performance, reliability and safety requirements. \ac{ctef} should replicate with high fidelity systems, services and capabilities of a \ac{cits} and provide the tools for investigating cyber threats in an ``\textit{in-vitro}'' or ``\textit{in-situ}'' fashion. These tools can be physical or virtual implementations that duplicate the real-world interactions between the different systems. 
    \item \textbf{\textit{Cybersecure Real-world Intelligent Framework}}: The knowledge acquired from the \ac{ctef}, should be easily transferable to the real-world. Of course, bidirectional interaction between the two entities is paramount, as identified cyber-threats in the real world should be replicated in \ac{ctef} and addressed with novel mitigation strategies in an ``in-vivo'' fashion. The real-time monitoring of systemic interactions can enhance the detection and prevention of risks. This framework should address traditional threats and consider behavioural aspects of \acp{cav} and \acp{cits}. The continuous operation of such a system, ensuring that disruptions will not affect the available services, is one of the key aspects that should be considered.
    \item \textbf{\textit{\ac{csce} Ecosystem}}: The above physical and virtual infrastructure solutions can address the operational aspects of such a platform. However, the surrounding ecosystem and an incident analysis framework within that is critical. National \ac{cits} systems' vulnerability, firmware version databases and bug bounty programs could provide the \ac{csce} with relevant tests and firmware version requirements to ensure all components will be continuously monitored and checked against a set of minimum requirements. Finally, maturity models can ensure high-quality solutions. Understanding the causes of the incident and providing feedback to the test facilities will allow fast prototyping and quick dissemination of solutions. Therefore, our \ac{csce} must be updated (through maturity cycles) in how it operates and accurately assesses future CAV systems.  
\end{enumerate}

The socio-technical aspects of \ac{csce} should be extended beyond technology, intertwining human, organisational and technological facets. Addressing cybercrime from other avenues (human-centric, educational-centric, etc.) can reduce its effects. Apart from the social benefits, this could also unlock new revenue opportunities and added economic value, thus making the business aspects part of the ecosystem. The five socio-technical areas to be considered are as follows:

\begin{itemize}[left=0em]
    \item \textbf{Human Factors}: Enhancing cybersecurity awareness is crucial. Training should focus on understanding risks, recognising vulnerabilities, and adopting protective best practices.
    \item \textbf{Public Engagement}: Trust in \acp{cits} and \acp{cav} is vital. Open dialogues can address concerns, dispel myths, and foster collective responsibility.
    \item \textbf{Organisational Dynamics}: Organisations should prioritise cybersecurity in strategies, policies, and operations. Regular audits and feedback loops ensure continuous improvement.
    \item \textbf{Collaborative Frameworks}: The complexity of \acp{cits} and \acp{cav} demands cross-sector collaboration. Shared initiatives can pool knowledge, resources, and best practices.
    \item \textbf{Regulatory and Policy Implications}: Evolving \acp{cits} and \ac{cav} landscapes require adaptive regulations. Policymakers should collaborate with experts to develop robust, flexible rules prioritising safety while encouraging innovation.
\end{itemize}

\subsection{The Different Dimensions of C-ITS Cybersecurity}
A holistic approach to cybersecurity, addressing both human and technical dimensions, ensures a resilient transportation ecosystem. For the remaining paper, the above entities will be further described, focusing primarily on the test facilities and their integration with the real world. Within \ac{csce}, and during the evaluation and testing of the different subsystems, different dimensions (layers) of cybersecurity should be considered for both \acp{cits} and fleets of \acp{cav}. These layers can be grouped as follows:
\begin{itemize}[left=0em]
    \item \textbf{Baseline Information Systems Risk:} A \ac{cav} is an information system that inherits the standard cyber risks and issues associated with connected information technologies.
    \item \textbf{Domain-specific Risk:} \ac{cav} systems provide domain-specific functions and services, each creating specialised attack surfaces that may affect cybersecurity posture.
    \item \textbf{Consequential Risk:} As cyber-physical systems, the operations of \ac{cav} platforms have consequences in the physical world, and these must be reflected in the assessment and appraisal of \ac{cav} systems.
    \item \textbf{Emergent Risk:} The autonomous component of these platforms introduces risks relating to emergentism -- that the system may autonomously arrive at and display behaviours not foreseen or intended by its creators.
\end{itemize}

\section{CSCE Architecture and System Design}\label{sec:systemDesign}
Secs.~\ref{sub:citsSystemDesign} and~\ref{subsec:requirements} discussed the requirement for a scalable, adaptable, modular, and extensible architecture. One pertinent paradigm is the \ac{oran} architecture~\cite{oranExplained}. \ac{oran} is built on open interfaces and emphasises its disaggregated and virtualised architecture. It promotes cloud-native applications that are modular, scalable, and easily upgradeable. We envision a similar architecture built upon microservices operating in a containerised cloud-native fashion. Such an architecture can facilitate our system's dynamic optimisation and integration with multi-dimensional communication planes. Moreover, the \ac{ctef} can harness diverse and distributed multi-vendor environments, ensuring the robustness and flexibility of its cybersecurity mechanisms while still maintaining interoperability across the different systems and sub-systems~\cite{cloudNative}. A distributed design can overcome bottlenecks, especially during high-demand periods~\cite{cloudNativeTesting}.

Overall, such an architecture will promote rapidly deploying highly extensible solutions easily transferable to the real world. This flexibility can enable integrating emerging technologies such as quantum computing, advanced AI algorithms, or new cybersecurity tools and tests. By providing finer-grained control over new features, we can speed up the development of new cybersecurity policies in the \ac{ctef} facilities, foster collaboration between research organisations, and provide a robust framework for research and innovation.

Two critical aspects of the system are its continuous learning and improvement and the focus on open standards and interoperability. Through real-time monitoring and feedback loops, \ac{csce} can continuously learn from its operations, adapt to new threats, and improve its defences, ensuring the system remains robust and relevant. By adhering to open standards, the \ac{csce} ensures seamless integration with other systems and provides a ``common language'' for various engineers.

Working in a \ac{secaas} fashion, the \ac{csce} can provide all the above-mentioned tools and services for automation, self-management and scalability in an as-a-service fashion. This will allow individuals and corporations to test their solutions without substantial capital outlays or complex initial implementations. When deployed in the real world, the solutions implemented can benefit from such an approach. The elasticity of these approaches and the benefit of short-lived  virtualised platforms can reduce the ``window of opportunity'' for attackers while maintaining availability, fast response times, disaster recovery and promoting vendor partnership and collaborations.

Finally, our \ac{ctef} and \ac{csce} will advocate for Digital Twin implementations, allowing cybersecurity solutions to be demonstrated using real-world data in virtual environments. This ensures that developed solutions will be directly applicable in the real world. A high-level conceptual diagram of this architecture can be seen in Fig.~\ref{fig:citsSystem}. The following sections will discuss the requirements for the testing facilities, the technologies that can facilitate the proposed \ac{ctef} and \ac{csce} and existing challenges.

\section{Designing a CTEF}\label{sec:designCTEF}
As explained earlier, the \ac{ctef} assesses the cybersecurity of software and hardware \ac{cits} components. It is essential to perform a comprehensive analysis that includes both ``in-vitro'' and ``in-situ'' testing to ensure fast sample analysis and improved system performance. In the upcoming sections, we will provide a detailed description of these two testing categories along with their corresponding requirements.

\subsection{CTEF: In-vitro  Testing and Analysis}\label{subsec:invitro}
In-vitro testing is an automated analysis of software samples outside their system context. It can identify malicious software through static and dynamic analysis in isolated virtual environments~\cite{practicalAnalysis}. In-vitro testing can minimise the overhead and resource utilisation of time-consuming penetration tests and systemic behavioural analyses and serves as the entry point for software samples to be tested in the \ac{ctef} or in parallel with in-situ tests.

In-vitro testing should meet traditional IT malware analysis requirements~\cite{malwareDetection,malwareDetection2}. The operational requirements that a \ac{cits} introduces are not considered at this stage since they are addressed in the in-situ tests. However, in-vitro tests should still provide efficient and scalable functionality to accommodate many concurrent tests. Sec.~\ref{subsub:inVitroHighLevel} provides the methodology for in-vitro testing, and Tab. \ref{tab:invitroRequirements} summarises the key functional requirements. More specifically, we have:

\begin{table}[t] 
\renewcommand{\arraystretch}{1.1}
\centering
    \caption{Requirements for In-vitro testing and analysis.}
    \begin{tabular}{lp{0.8\columnwidth}}
    \toprule
    \textbf{Req.} & \textbf{Requirement Description} \\ \midrule
    \textbf{VTR1} & Must be able to detect existing malware \\
    \textbf{VTR2} & Must be able to detect unknown malware \\
    \textbf{VTR3} & Must be able to detect software vulnerabilities \\
    \textbf{VTR4} & Must consider and defend against attacks to the detection mechanisms \\
    \textbf{VTR5} & Must be scalable to allow concurrent testing and analysis of multiple CAV software samples \\
    \textbf{VTR6} & Must enable an easy transition to a clear state \\
    \textbf{VTR7} & Must be updatable on the fly to incorporate new cyber threat detection capabilities \\
    \textbf{VTR8} & Must be isolated from the outside world to avoid network contamination \\
    \textbf{VTR9} & Must be able to provide multiple OS environments for testing and analysis \\
    \textbf{VTR10} & Must maintain a log of scanned software and produce relevant reports \\
    \textbf{VTR11} & Should allow a user to create an account and test their software \\
    \textbf{VTR12} & Software samples that are to be tested should not be visible to anyone apart from the user that submitted the software \\
    \bottomrule
    \end{tabular}
    \label{tab:invitroRequirements}
\end{table}

\begin{enumerate}[wide, labelwidth=!, labelindent=9pt, start=1,label={\bfseries VTR\arabic*}]
    \item \textbf{-- Detection of known malware}: Malicious software is identified by evaluating specific instructions and/or byte sequences against known vulnerability databases. An efficient automated signature generation method should be implemented, as described in \cite{10.1007/978-3-642-04342-0_6}, to reduce the time required for static analysis, also linked to a requirement for continuous integration (Req.~\ref{req:vtrreq7}). \label{req:vtrreq1}
    \item \textbf{-- Detection of unknown malware}: Zero-day attacks, i.e., malware that exploits vulnerabilities not known before and malware that transforms their code to evade signature-based detection mechanisms, must be detectable within \ac{ctef}, rendering traditional signature-based malware detection tools as insufficient. \ac{ml} strategies could be employed to detect such malicious behaviours~\cite{unknownMalware}. Potential datasets for that are NSL-KDD~\cite{kddDataset} and its extensions, consisting of 125k samples and 41 features, EMBER~\cite{anderson2018ember}, with more than 1M samples, and more. The detection capability can be based on detecting anomalies in the normal operation of the virtual environment when the software sample is executed in it (anomaly detection using unsupervised learning) or on the similarities in the behaviour of the software sample with the behaviour of known malware (supervised learning). \label{req:vtrreq2}
    \item \textbf{-- Software vulnerabilities detection}: Bad coding practices and security holes in the software can result in vulnerabilities. Other malicious software and malevolent actors can exploit these code vulnerabilities. In the \ac{ctef}, vulnerability detection will be realised in two ways: \label{req:vtrreq3}
    \begin{itemize}[left=0.5em]
        \item \ac{ml} techniques can be used to automate the detection of vulnerabilities in the source code~\cite{vulnerabilityDetectionSource}. The existence of a large open source codebase favours the training of \ac{ml} (e.g., multiclass classification of source code vulnerabilities using deep learning or Recurrent Neural Networks when only binaries are available).
        \item Where the source code is unavailable, \ac{ml} techniques will be used to automate the detection of vulnerabilities in the binary code~\cite{vulnerabilityDetectionBinary}.
    \end{itemize}
    \item \textbf{-- Defence against adversarial \ac{ml}}: \ac{ml} can detect unknown malware and vulnerabilities, but it is vulnerable to attacks such as input manipulation, model attack and model theft that aim to force deliberate misclassification of inputs~\cite{adversarialML}. \ac{ctef} must consider defence strategies against these attacks and also integrate new solutions into the implemented pipelines, leveraging its extensibility and scalability (Req.~\ref{req:vtrreq5}). \label{req:vtrreq4}
    
\begin{figure*}[t]     
\centering
\includegraphics[width=1\textwidth]{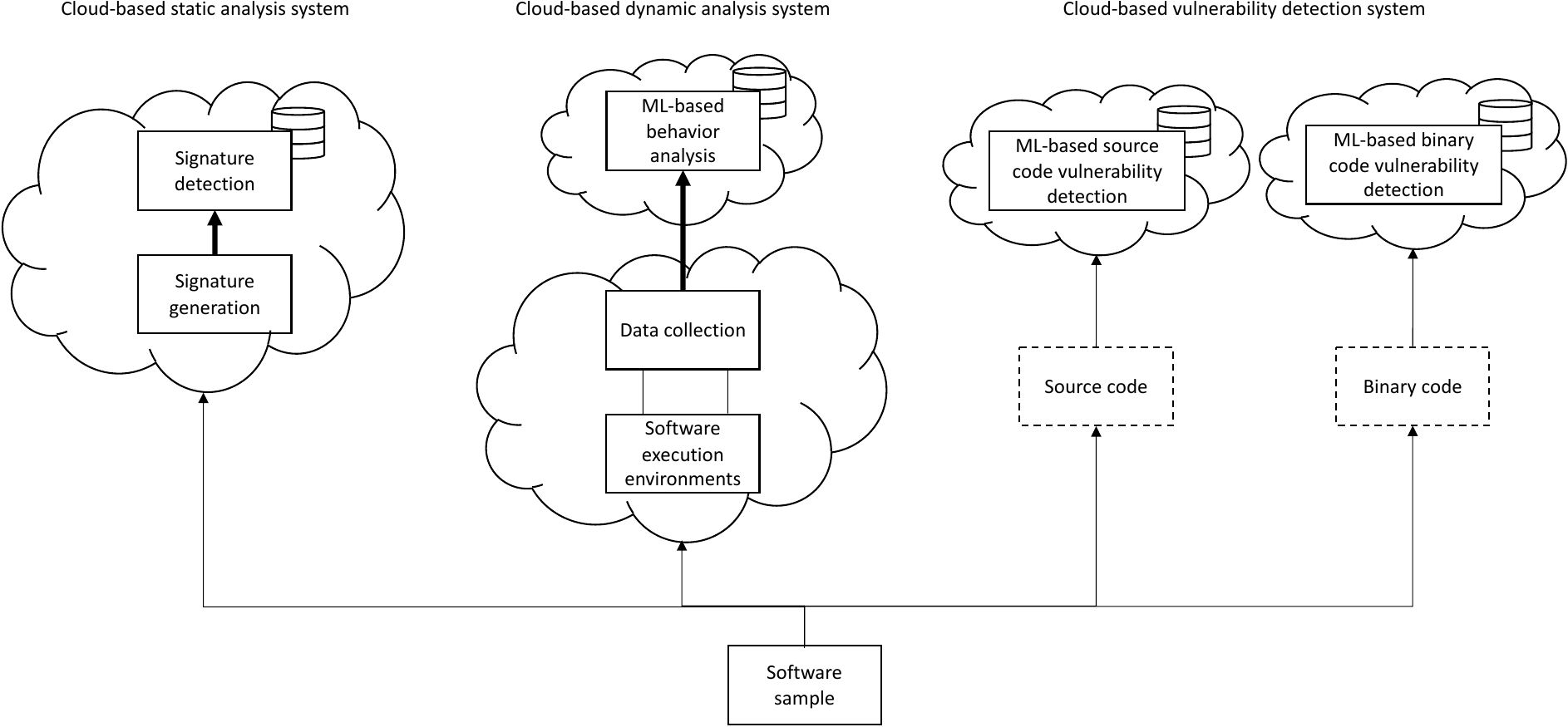}
\caption{Example of an in-vitro's cloud-based software analysis system.}
\label{fig:iv_software}
\end{figure*}
    
    \item \textbf{-- Scalability}: In-vitro testing will use cloud-native services for scalability and concurrent testing. Tests will run in ephemeral and isolated environments (e.g., in the cloud or containers)~\cite{cloudNativeTesting}, as presented in Figure \ref{fig:iv_software}. The environments must reset to their initial state after each test. \label{req:vtrreq5}
    \item \textbf{-- Easy restoration to a clean state}: Similar to Req.~\ref{req:vtrreq5}, virtualised environments will enable easy reset and restoration of the system to a clean state after the end of a test~\cite{cloudNativeTesting}. This, in conjunction with multiple virtual environments, will allow for more efficient operation, improving the overall performance of the \ac{ctef}. \label{req:vtrreq6}
    \item \textbf{-- Update on the fly}: In-vitro tests must stay updated with new cyber threats. \ac{ml}-based detection models must be regularly retrained to avoid concept drift over time~\cite{conceptDrift} and improve their detection capability. Additionally, adopting multiple well-supported open-source signature-based malware detection tools will further improve the detection effectiveness. Finally, virtualised environments can enable real-time updates and integration of new detection capabilities. \label{req:vtrreq7}
    \item \textbf{-- Avoid network contamination}: Using virtualised testing environments will enable the isolation of the in-vitro tests from the outside world, thus avoiding network contamination. The virtual environments must be appropriately set up to avoid external network connections. \label{req:vtrreq8}
    \item \textbf{-- Multiple OSs}: Similar to Req.~\ref{req:vtrreq8}, the use of virtualised environments will enable the running of a variety of operating systems, which in turn will allow software execution and behaviour analysis in the appropriate set-up. \label{req:vtrreq9}
    \item \textbf{-- Logging and reporting}: Each test in the in-vitro environment will be logged along with the test output in a centralised database. The database will keep a record of all users and software tested. The \ac{ctef} must follow traditional IT systems-based information risk management and privacy policies and regulations (e.g. ISO 27001~\cite{ISO27001}, \ac{gdpr}, etc.) to ensure appropriate data privacy and protection. \label{req:vtrreq10}
    \item \textbf{- User interaction}: The CTEF should provide a user interface to allow users to upload their software for testing. Similarly to Req.~\ref{req:vtrreq10}, the user interfaces and backend implementations should adhere to best practices and standards for IT systems-based information risk management (e.g. ISO 27001~\cite{ISO27001}, \ac{gdpr}, etc.). \label{req:vtrreq11}
    \item \textbf{-- Manufacturers Confidentiality}: To ensure manufacturers' confidentiality, software samples for testing will only be accessible to the submitting user. Virtual environments used in testing must not be accessible by other users and will be reset and restored after each test. Data stored in the \ac{ctef} must be visible only to the user whose software generated the data. \label{req:vtrreq12}
\end{enumerate}

\subsection{High-Level Design for In-vitro tests}\label{subsub:inVitroHighLevel}

\begin{figure}[t]     
\centering
\includegraphics[width=1\columnwidth]{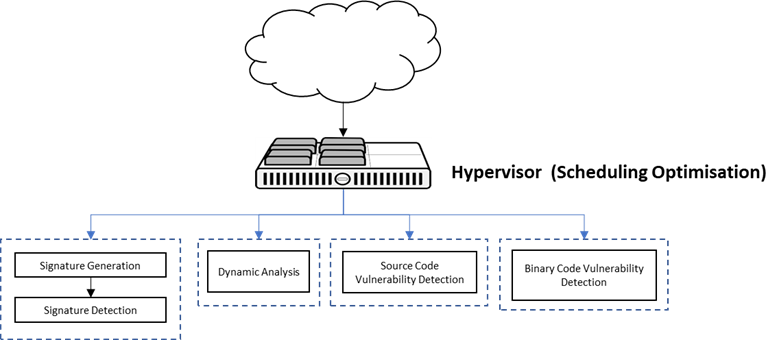}
    \caption{In-vitro \ac{ctef} testing and optimisation regime.}
    \label{fig:inVitro}
\end{figure}

The in-vitro testing is broken down into four subtests: static analysis, dynamic (behavioural) analysis, vulnerability detection in the source code and vulnerability detection in the binary code as shown in Fig.~\ref{fig:inVitro}. Each subtest will be executed in an isolated virtual environment addressing the Reqs.~\ref{req:vtrreq5} to~\ref{req:vtrreq9}.

Scheduling optimisation can be performed by a ``hypervisor/orchestrator'' -(Fig.~\ref{fig:inVitro}). After submitting the software samples to be tested, the hypervisor can determine the number of virtual environments needed and the scheduling of subtests. We expect the number of virtual environments needed to be dynamically changed so that software samples that fail one of the subtests do not consume system resources. The complexity and performance of each subtest will also be considered in the scheduling process. For instance, static analysis is much faster and resource-light than dynamic analysis; as such, in a single sample analysis case, static analysis should always precede dynamic analysis. Resource allocation/deallocation and subtask scheduling will be automated to further improve the testing system’s performance. 

\begin{figure}[t]     
\centering
\includegraphics[width=1\columnwidth]{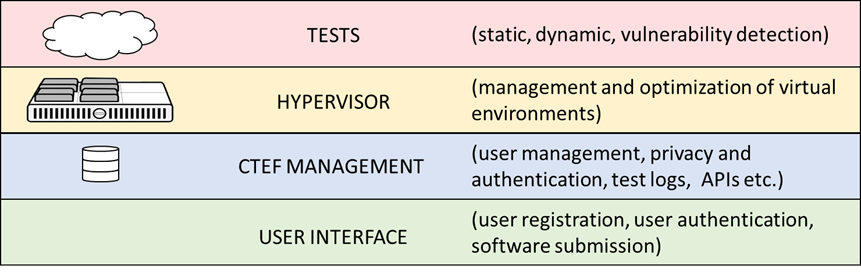}
    \caption{In-vitro high-level architecture diagram.}
    \label{fig:highLevelInVitro}
\end{figure}

So, based on the above, we envision an in-vitro environment that will consist of four components (Fig.~\ref{fig:highLevelInVitro}): 
\begin{enumerate}[left=0.5em]
    \item the user interface allowing user registration and software submission to the system in a secure way,
    \item the \ac{ctef} management and configuration system providing the appropriate APIs to run the desired tests and user management, test configuration and logging capabilities, 
    \item the hypervisor for spawning virtual tests and optimise the use of resources in a cloud-native manner, and finally,
    \item the actual tests that will be run in the spawned virtual environments. 
\end{enumerate}

\subsection{CTEF: In-situ Testing and Analysis}\label{subsec:insitu}
In-vitro testing can detect malicious software and vulnerabilities but does not account for the interaction between \ac{cav} and \ac{cits} infrastructure. In-situ tests allow \ac{cav} systems to interact with virtual \acp{cav} and \ac{cits}-related environments, providing a more realistic evaluation of software and hardware components. For example, a malicious \ac{cav} may aim to brake suddenly at a busy intersection to block or collide with other vehicles. While triggering the \ac{cav} braking control sequence may not seem suspicious or malicious on its own, the context of where the \ac{cav} stopped, i.e., in the middle of a busy intersection, suggests malicious behaviour. In-situ testing aims to fill this gap by providing realistic \ac{cits} environments for testing and evaluation.

\begin{figure}[t]     
\centering
\includegraphics[width=1\columnwidth]{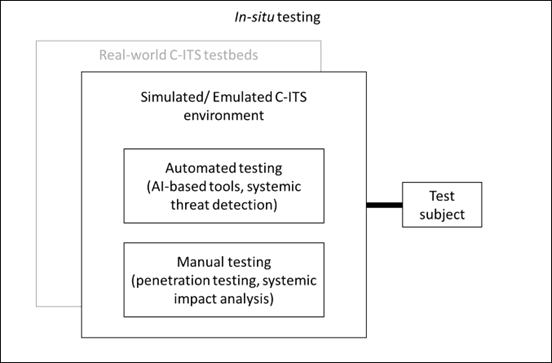}
    \caption{Proposed in-situ testing scheme.}
    \label{fig:inSituTestingScheme}
\end{figure}

Rigorous, transparent, and replicable cybersecurity in-situ testing of new hardware and software will occur in simulated \ac{cits} environments in \ac{cits} testbeds. Testing environments will evaluate the security of new technologies and analyse their impact on the system’s operations and performance. These tests will not replace traditional penetration testing techniques. On the contrary, in-situ tests will be operated as automated standalone tests in conjunction with penetration testers while emulating real-world \ac{cits} scenarios (Fig.~\ref{fig:inSituTestingScheme}). In such a setup, penetration tests will investigate the \ac{v2x} data flows within an \ac{cits}. The collected data will be analysed using \ac{ml}-based approaches to detect potential systemic malicious behaviours. Sec.~\ref{subsub:inSituHighLevel} provides the methodology for in-vitro testing and analysis, and Tab.~\ref{tab:inSituRequirements} summarises its requirements. The requirements are tailored to the needs of a \ac{cits} and account for the lessons learned from previous cyber-security testbeds~\cite{previousCybersecTestbeds}. Briefly, we have: 

\begin{table}[t] 
\renewcommand{\arraystretch}{1.1}
\centering
    \caption{Requirements for In-situ testing and analysis.}
    \begin{tabular}{l p{0.8\columnwidth}}
    \toprule
    \textbf{Req.} & \textbf{Requirement Description} \\ \midrule
    \textbf{ST1} & Must be run in a configurable \ac{cits} test environment \\
    \textbf{ST2} & Must support multiple \ac{cits} configurations \\
    \textbf{ST3} & Must support simulation of processes and devices \\
    \textbf{ST4} & Must support all data flows identified in \ac{cits} \\
    \textbf{ST5} & Must include a diverse range of devices and protocols \\
    \textbf{ST6} & Must produce performance metrics based on the \ac{cits} operational requirements \\
    \textbf{ST7} & Must detect vulnerabilities by performing penetration tests \\
    \textbf{ST8} & Must run a series of cyber attack scenarios \\
    \textbf{ST9} & Must provide tools to perform penetration tests \\
    \textbf{ST10} & Must detect links between cyber-attack scenarios and misuse scenarios \\
    \textbf{ST11} & Must provide data logging capability for all data flows \\
    \textbf{ST12} & Must provide the tools to perform behaviour analysis for threat detection \\
    \textbf{ST13} & Must enable an easy transition to the initial/clean state \\
    \textbf{ST14} & Must keep the complexity of experimental infrastructure at check \\
    \textbf{ST15} & Must be designed with scalability and flexibility in mind in order to be transferable to the real world \\ \bottomrule
    \end{tabular}
    \label{tab:inSituRequirements}
\end{table}

\begin{figure}[t]     
\centering
\includegraphics[width=1\columnwidth]{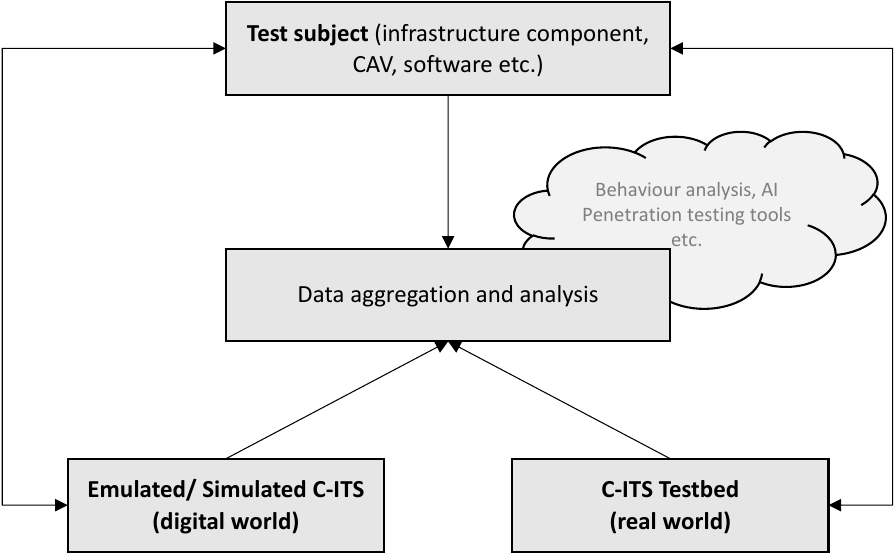}
    \caption{In-situ high-level interconnections.}
    \label{fig:is_hl}
\end{figure}

\begin{enumerate}[wide, labelwidth=!, labelindent=9pt, start=1,label={\bfseries ST\arabic*}]
    \item \textbf{-- Configurable \ac{cits} test environment}: In-situ tests should run in simulated/emulated \ac{cits} environments, either digital (a computerised simulation of a \ac{cits}) or real-world (e.g., Zenzic’s testbeds\footnote{Zenzic \ac{cits} Testing in the UK: https://zenzic.io/testbed-uk/}). In-situ testing could be based on the packet-level vehicular simulation frameworks such as Veins~\cite{veins} integrated with traffic simulators such as SUMO~\cite{SUMO}. This will allow for the monitoring and analysis of communication patterns and the detection of network anomalies. \ac{cav} or \ac{cits} related hardware and software could be ``plugged in'' to the simulated environment and interact with digital entities (simulated vehicles and/or infrastructure) at the network level (Fig. \ref{fig:is_hl}). \label{req:streq1}
    \item \textbf{-- Support multiple \ac{cits} configurations}: The use of a digitally simulated \ac{cits} environment will enable the deployment of a variety of different scenarios with different system configurations. These configurations must align with the requirements of real-world \ac{cits} testbeds to ensure rapid development, testing and deployment. \label{req:streq2}
    \item \textbf{-- Simulation of devices or processes}: The simulated \ac{cits} environment can enable the simulation of devices and processes and create more realistic and complex scenarios. Exposing various APIs can enable an easier connection between the tested software and the simulation framework. \label{req:streq3}
    \item \textbf{-- Support all data flows in \ac{cits}}: To better mimic the real world, the simulator/emulator used must support all data flows identified in a \ac{cits} (as described in Secs.~\ref{subsec:attackSurface} and~\ref{subsec:modellingData} and shown in Fig.~\ref{fig:referenceArchitecture}). \label{req:streq4}
    \item \textbf{-- Diverse devices and protocols}: The \ac{cits} and vehicular communication protocols (e.g., IEEE ITS-G5, \ac{3gpp} LTE, etc.) should be supported by the simulation frameworks, implementing the different standards. Additionally, cybersecurity tools used to evaluate the security of the software/hardware undergoing testing must be compatible with these communication protocols. \label{req:streq5}
    \item \textbf{-- Performance metrics}: A baseline profile should represent the system's behaviour without adding other software or hardware. This profile, generated before any modification, can be compared to the network performance achieved (jitter, throughput, etc.) after adding the new software or hardware to the testing environment. \label{req:streq6}
    \item \textbf{-- Vulnerabilities detection/Penetration testing}: Penetration tests can provide insight into the likelihood of a threat occurring and how successful a cyber attack could be\footnote{NCSC penetration guide for testing: https://www.ncsc.gov.uk/ guidance/penetration-testing}. Such capabilities must be incorporated in in-situ testing. Following the guidelines on cybersecurity provided by the US Department of Transportation~\cite{citsGuidelines}, the scope of the penetration tests should include security policies, devices, applications, networks, access controls, communications and configurations that can compromise the \ac{cits}. In-situ penetration tests will consider the following: \label{req:streq7}
    \begin{itemize}[left=0.5em]
        \item \textbf{Infrastructure}: Includes field devices such as traffic sensors, traffic control and signalling, public messaging etc. and the wireless and wired networks that support them.
        \item \textbf{Traveller}: Encompasses the devices used by the traveller to access \ac{cits} services (e.g. traffic or emergency notifications).
        \item \textbf{\acp{cav}}: Software and hardware found in \acp{cav} and communication with other entities in a \ac{cits}.
        \item \textbf{Communications}: Includes the communication components of the \ac{cits}. These include various wireless technologies such as Wi-Fi, WiMAX, mmWAVE, cellular networking, etc.
    \end{itemize}
    Finally, tests should be provided on all categories, i.e., black, grey and white box testing \cite{penTestCategories}. These tests should be structured based on the attack scenarios we describe in Req.~\ref{req:streq8} and cover all phases of a cyber attack, as described by the Cyber Kill Chain Model~\cite{killChain}.
    \item \textbf{-- Cyber attack scenarios}: Penetration tests must be structured on a series of cyber attack scenarios tailored to \ac{cits}, as discussed in~\cite{citsAttacks,citsAttacks2}. The selection of the attack scenarios is controlled by the penetration tester, depending on the component to be tested. Attack scenarios can be categorised into Physical, Wireless, Network and \acp{vanet}. Tab.~\ref{tab:attacks} summarises the attack scenarios, excluding organisation attacks, against components in a \ac{cits}. More details about these attacks can be found in~\cite{attacks1,attacks2}. \label{req:streq8}
    \item \textbf{-- Penetration testing tools}: To enable the execution of the attack scenarios described in Req.~\ref{req:streq8}, in-situ will also provide the appropriate penetration testing tools described by the Penetration Testing Execution Standard~\cite{penetrationStandards}. The testing tools will be integrated into a penetration testing suite and executed in a cloud-native-based approach inside a virtualised environment. Tools must be compatible with all technologies and communication protocols in \ac{cits}. \label{req:streq9}
    \item \textbf{-- Link cyber attacks to misuse scenarios}: To allow for better threat analysis, in-situ testing will link the attack scenarios presented in Tab.~\ref{tab:attacks}. \label{req:streq10}

\begin{table*}[!htbp]
\renewcommand{\arraystretch}{1}
\caption{Cyber attacks against \ac{cits}.}
\centering
\begin{tabularx}{\linewidth}{>{\hsize=.4\hsize\linewidth=\hsize}X>{\hsize=1.5\hsize\linewidth=\hsize}X}
\toprule
\textbf{Attack Types}         & \textbf{Attacks}\\ 
\midrule
    \multirow{13}{\linewidth}{\textbf{Physical Attacks} (attacks deployed by having physical access to the component to be tested)} &   \tabitem Physically connecting to exposed ports, e.g., USB, serial, etc.  \\
    & \tabitem Using brute force or guessing credentials on a device  \\
    & \tabitem Sniffing network traffic between a device and the backend \\
    & \tabitem Scanning the secured/closed network to discover the topology \\
    & \tabitem Deleting files on the compromised \ac{cits} device/system \\
    & \tabitem Dumping firmware to recover credentials and configuration \\
    & \tabitem \ac{mitm} attacks using exposed wires/cables to intercept data \\
    & \tabitem Physically tamper with a device to steal/compromise data, modifying a device, etc. \\ 
    & \tabitem Connecting a removable storage device loaded with malware to install \\  
    & \tabitem Sending improper commands to the controller and backend servers \\
    & \tabitem \ac{mitm} attack communications and sending false data to backend servers \\
    & \tabitem Pivoting a \ac{cits} device as a trusted entry point into the corporate network \\
    & \tabitem Exploiting vulnerabilities in software, hardware, protocols, OS, etc. \\  \midrule
    
    \multirow{12}{\linewidth}{\textbf{Wireless Attacks} (attacks deployed against or/and via wireless communications)} &   \tabitem Spoofing V2V, V2I, and I2I messages broadcast to traffic and the rest of the \ac{cits} ecosystem  \\
    & \tabitem Sniffing wireless transmissions, e.g., using the car’s Wi-Fi  \\
    & \tabitem Remotely transmitting and installing malicious firmware \\
    & \tabitem Electronic jamming of wireless transmissions to disrupt operations \\
    & \tabitem \ac{mitm} attack with wireless transmission to intercept and/or modify data \\
    & \tabitem Exploiting vulnerabilities in software, hardware, protocols, OS, etc. \\
    & \tabitem Using vehicle Wi-Fi as an entry point into the controller area network (CAN) bus and then to the on-board diagnostics (OBD), telematics control unit (TCU), and in-vehicle infotainment (IVI) \\
    & \tabitem The remote hijacking of vehicle controls via compromised CAN bus  \\
    & \tabitem Installing malicious third-party apps in a car’s infotainment system \\  
    & \tabitem Attacking via a malicious app installed on a phone connected to the car’s Wi-Fi \\
    & \tabitem Electronic jamming of vehicle’s safety systems, e.g., radar, ultrasonic sensors, etc. \\ 
    \midrule
    \multirow{15}{\linewidth}{\textbf{Network Attacks} (attacks that take advantage of the connection of \ac{cits} components to the Internet (Internet-exposed \ac{cits} systems)} &   \tabitem Identifying and abusing device misconfigurations  \\
    & \tabitem Exploiting vulnerabilities in legacy software and hardware  \\
    & \tabitem Installing malware/spyware on systems \\
    & \tabitem Uploading and installing malicious firmware \\
    & \tabitem Launching D\ac{dos} attacks on internet exposed \ac{cits} infrastructure and backend servers \\
    & \tabitem Exploiting vulnerabilities in software, hardware, protocols, OS, etc. \\
    & \tabitem Credential brute-forcing and abusing weak authentication mechanisms \\
    & \tabitem SQL injection attacks  \\
    & \tabitem Cross-site scripting (XSS) attacks \\  
    & \tabitem Session hijacking attacks \\
    & \tabitem DNS spoofing and hijacking attacks \\ 
    & \tabitem Pass the Hash attacks \\ 
    & \tabitem Pass the Ticket attacks (Kerberos) \\
    & \tabitem Sending improper commands to the controller and backend servers \\ 
    & \tabitem Pivoting an ITS device as a trusted entry point into the corporate network \\ 
    \midrule 
    \multirow{7}{\linewidth}{\textbf{Attacks against VANETs} (attacks exploiting Vehicular Ad Hoc Networks)} &   \tabitem Sybil  \\
    & \tabitem D\ac{dos}  \\
    & \tabitem Blackhole \\
    & \tabitem Wormhole \\
    & \tabitem False information \\
    & \tabitem Replay \\
    & \tabitem Passive eavesdropping \\ \bottomrule 
\end{tabularx}
\label{tab:attacks}
\end{table*}

    \item \textbf{-- Data flow collection and storage}: To enable better analysis and aid for research and development, all data flows must be stored anonymously and securely. Furthermore, the tests and actions should be logged and aligned with the stored data. Data stored can aid in researching and developing new cybersecurity tools and testing methods. \label{req:streq11}
    \item \textbf{-- \ac{ml}-based threat detection}: Appropriate \ac{ml}-based threat detection algorithms should be used to detect abnormal behaviour within the simulated \ac{cits} environment. Threat detection must be performed before and after penetration tests to ensure the detection of malicious changes in the system that exploit vulnerabilities. This will allow for the detection of malicious hardware/software as well as vulnerability exploitation, as highlighted in Fig.~\ref{fig:inSituThreat}. \label{req:streq12}
    \item \textbf{-- Easy transition to initial state}: Virtualised environment must return to an initial state after testing. By doing so, we can ensure that controlled environments are always used for experimentation. \label{req:streq13}
    \item \textbf{-- Managing the complexity}: A digital network management system should be deployed to oversee the operation of the \ac{ctef} as tests are being carried out. The management system should be responsible for the following operations: \label{req:streq14}
    \begin{itemize}[left=0.5em]
        \item Register new tests.
        \item Allow the configuration of a new \ac{cits} testing environment (in case of a digital simulation) or connect to real-world \acp{cits} testing environments.
        \item Spawn new \ac{cits} simulation environments (in case of a digital simulation). The tested SW/HW will be ``plugged '' into the simulation environment.
        \item Capture, store, and visualise all generated data within the \ac{cits} (digital or real-world) testing environment. Present system performance metrics.
        \item Analyse the generated data using cloud-based \ac{ml}-driven cybersecurity tools (aligns with Req.~\ref{req:streq12}).
        \item Initiate cloud-based penetration testing platform to be used by the facility’s penetration testers (aligns with Reqs.~\ref{req:streq7},~\ref{req:streq8} and~\ref{req:streq9}).
        \item Log executed tests and results.
        \item Update the data analysis tools and the penetration testing platform to address new threat vectors.
        \item Terminate running tests.
    \end{itemize}
    \ac{ctef} must follow traditional IT systems-based information risk management and privacy policies and regulations (e.g. ISO 27001~\cite{ISO27001}, \ac{gdpr}, etc.) to ensure user data privacy and cyber protection. More information is provided in Sec.~\ref{sub:securityObjectives}.
    \item \textbf{-- Transferability to the real world}: In-situ tests and solutions must be designed with scalability and flexibility in mind. This allows easy integration into the real world. The adoption of a \ac{secaas} approach, along with the use of simulated \ac{cits} environments, satisfy the two requirements. Using a single management network to which all tools and testing environments are connected favours the extensibility. \label{req:streq15}
\end{enumerate}

\begin{figure}[t]     
\centering
\includegraphics[width=0.85\columnwidth]{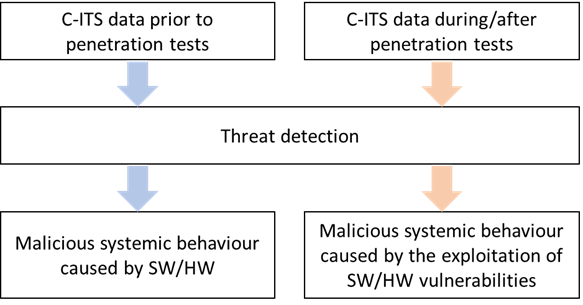}
    \caption{In-situ \ac{cits} threat detection.}
    \label{fig:inSituThreat}
\end{figure}

\subsection{High-Level Design for In-situ tests}\label{subsub:inSituHighLevel}
\begin{figure}[t]     
\centering
\includegraphics[width=1\columnwidth]{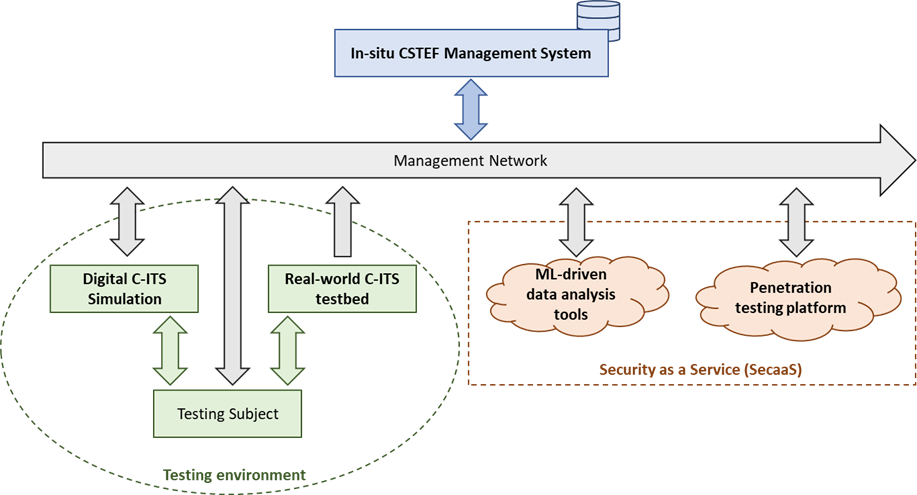}
    \caption{In-situ high-level architecture diagram.}
    \label{fig:highLevelInSitu}
\end{figure}

In-situ \ac{ctef} is broken down into five stages as presented in Fig.~\ref{fig:highLevelInSitu}. This includes the Management System, the Digital \ac{cits} simulation, the Real-world \ac{cits} testbed, the \ac{ml}-driven data analysis tools and the penetration testing platform. 

The data analysis tools and penetration testing platform will run cloud-natively. Penetration testers can use the platform to run the attack scenarios described in Req.~\ref{req:streq8}. The penetration testing platform can be implemented independently of the other components. The ML data analysis tools depend on data from the digital and real-world \ac{cits} simulations. ML algorithms will be trained with simulator data and then improved with real-world data.

The \ac{cits} simulation environment can be cloud-based or locally implemented. Cloud-based simulations offer flexibility but add latency. Local implementation reduces latency and favours flexibility and scalability while better representing a \ac{cits} environment. The management network connects \ac{secaas}, testing environments, and the \ac{ctef} Management System. Real-world testbeds will also be used, and existing ones will be adapted to grant \ac{secaas} platforms access to data planes.

\section{Technical \& Functional Specification}\label{sec:technologies}
This section describes the technologies that will provide \ac{ctef} with the modularity, extensibility, scalability and robustness required. We align each technology's benefits with the requirements described in Sec.~\ref{subsec:requirements}. As known, each state-of-the-art technology solves many issues but also brings new challenges, so in the following sections, we will delve into the cybersecurity challenges that need to be addressed for all the presented technologies.

\subsection{Microservices Architecture in Cybersecurity}
A microservice architecture~\cite{microservices} is a set of services loosely coupled to implement an application. These services provide finer-grained control over an entire application that operates as a whole. Microservices are rapidly gaining popularity in the \ac{sdlc} community as they facilitate continuous delivery for larger applications and more straightforward adaptation when the requirements are updated. The benefits with regards to cybersecurity and \ac{csce} are summarised as follows:

\begin{itemize}[left=0.5em]

    \item \textbf{Granular Security Controls}: As each microservice handles a specific function (e.g., TLS certificate dissemination, encryption of persistent data, etc.), security measures can be applied with more precision. This granularity facilitates independent scaling within a \ac{sos} to meet varying security demands and counteract specific threats (aligns with Reqs.~\ref{req:cscereq3}, ~\ref{req:cscereq8}, ~\ref{req:ctefreq1}, and~\ref{req:ctefreq2}).

    \item \textbf{Diverse Defense Mechanisms}: The flexibility of microservices allows each service to be developed using different languages or frameworks. This diversity makes it harder for attackers to exploit a systemic vulnerability, as each service could potentially present a unique defensive profile (aligns with Reqs.~\ref{req:cscereq3}, and~\ref{req:ctefreq1}).

    \item \textbf{Isolated Incident Management}: With modular design, breaches or vulnerabilities in one service do not necessarily compromise others. Isolation enables efficient troubleshooting and targeted fixes, enhancing the robustness of the overall system. Testing processes are streamlined since issues can be narrowed down quickly (aligns with Reqs.~\ref{req:cscereq1} and~\ref{req:ctefreq1}).

    \item \textbf{Secure Continuous Delivery}: Implementing security measures in microservices ensures that changes across the application's lifecycle are introduced without compromising security. When cross-functional teams, including developers, operations, and testing teams, focus on a single service, integrating security measures becomes an intrinsic part of the delivery process (aligns with Reqs.~\ref{req:cscereq5},~\ref{req:cscereq6},~\ref{req:ctefreq6}, and~\ref{req:ctefreq7}).

    \item \textbf{Proactive Threat Detection}: Continuous and robust monitoring ensures real-time oversight of the security landscape. In case of security anomalies or a service malfunction, predefined mitigation strategies can immediately spring into action, be it to restart a compromised service or initiate countermeasures against potential threats (aligns with Req.~\ref{req:cscereq6}).

\end{itemize}

\subsection{Cloud-native Architectures for Cybersecurity}\label{subsec:cloudNative}
A cloud-native infrastructure serves as the backbone to support the microservices approach vital for use-cases like \ac{cits} and Smart Cities~\cite{smartCities}. This approach is widespread in many Smart City, Robotics, and Infrastructure projects (e.g., as in~\cite{umbrellaRobot,smartCity}) and aligns with the key concepts for future \acp{cits}~\cite{mainPlanes}. Cloud-native architecture harnesses the inherent capabilities of cloud computing to enhance the security and efficiency of systems. Key benefits include automation of microservices, horizontal and vertical scaling, and resilience. This enables easier management and operations of complex microservices-based systems~\cite{cloudNative}.

Smart Cities and \acp{cits} are both inherently exposed to cyber threats due to their wide digital footprint. A cloud-native approach offers:

\begin{itemize}[left=0.5em]
    \item \textbf{Secure Automated Deployment}: Cloud-native ensures that deployment and management of numerous microservices are automated, reducing human error and potential security vulnerabilities (aligns with Reqs.\ref{req:ctefreq1} and\ref{req:ctefreq5}).
    \item \textbf{Dynamic Scaling}: As systems experience varying demand, cloud-native offers adaptive scaling without compromising security standards (aligns with Req.\ref{req:ctefreq1}).
    \item \textbf{Resilience}: By leveraging distributed architectures, resilience is assured even in the face of concentrated cyber attacks (aligns with Req.~\ref{req:cscereq8}).
    \item \textbf{Managed Security Services}: Cloud-native encourages the use of managed cloud services, like secure databases, encrypted messaging, and robust storage solutions, enhancing the security ecosystem (aligns with Req.~\ref{req:ctefreq2}).
\end{itemize}

This architectural advantage ensures that the \ac{ctef} and its subsystems remain agile, responsive, and secure against evolving threats. Cloud-native infrastructure becomes pivotal when meeting the \ac{secaas} specifications mentioned in \ref{req:ctefreq4}. By integrating \ac{saas} principles, security services are not only provisioned on-demand but also fortified and elastic. Overall, a well-designed cloud-native architecture will provide the foundation for \ac{ctef} and the extensibility and adaptability required for such a system.

\subsection{Containerisation of Software and Orchestration}
Containers are not just resource-efficient alternatives to traditional virtual machines but also enhance security by isolating application processes.  Containerisation Software (e.g., Docker\footnote{Docker: Platform-as-a-service for OS-level virtualisation: \url{https://www.docker.com/}}, LXC\footnote{LXC: Userspace interface for Linux Kernels: \url{https://linuxcontainers.org/lxc/}}, etc.) are cloud-native ecosystems that encapsulate applications in containerised environments, abstracting them from potential vulnerabilities of the underlying machinery. They are beneficial for the following reasons:

\begin{itemize}[left=0em]
    \item \textbf{Scalability}: Containers facilitate rapid, secure scaling from testing to deployment, ensuring no compromised components scale along (aligns with Req.~\ref{req:ctefreq1}).
    \item \textbf{Resilient}: The inherent resilience of containers ensures that potential breaches do not persist, as containers can self-recover, minimising exposure (aligns with Req.~\ref{req:cscereq8}).
    \item \textbf{Distributed}: Their distributed nature not only aids in resource management but also decentralises potential attack vectors, reducing single points of failure (aligns with Req.~\ref{req:ctefreq2}).
    \item \textbf{Portability}: Enables the application to run on various locations, i.e., on-premises, in a public cloud, or in a private cloud. Its independence from the host OS minimises the exposure to potential vulnerabilities there (aligns with Req.~\ref{req:cscereq3}).
\end{itemize}

For future \ac{cits}, a container orchestration layer that emphasises security is crucial~\cite{containerOrchestration}. Container orchestration is used to manage the lifecycle of containers, especially in large, dynamic environments. Container orchestration can be used to control and automate many tasks, such as:

\begin{itemize}[left=0em]
    \item Provisioning and deployment of containers (Req.~\ref{req:cscereq3}).
    \item Redundancy and availability of containers (Req.~\ref{req:cscereq8}).
    \item Scaling up or removing containers to spread application load evenly across host infrastructure
    \item Moving and rescheduling containers from one host to another. This ensures high availability of the services even when one or many hosts are offline or malfunctioning (Req.~\ref{req:ctefreq5}).
    \item External exposure of services running in a container with the outside world (Reqs.~\ref{req:cscereq3} and~\ref{req:ctefreq6}).
    \item Load balancing and service discovery between containers. This can help find the available services within the virtual network (Req.~\ref{req:ctefreq5}).
    \item Monitoring the health of the containerised applications and the host infrastructure (Req.~\ref{req:cscereq3}).
\end{itemize}

Such a flexible architecture can significantly benefit the massive growth of heterogeneous devices connected to such networks. The most popular tool for container orchestration currently is Kubernetes\footnote{Kubernetes: Production-grade container orchestration: \url{https://kubernetes.io/}}. Kubernetes is an open-source production-grade container-orchestration system for automating application deployment, scaling, and management. 

The benefits of container orchestration have been investigated in various Smart City and \ac{cits} deployments, e.g., in~\cite{KubeHICE,Böhm_Wirtz_2022}. These works show that benefits are related to these environments' delay-sensitive and data-intensive services. These requirements are crucial for such systems. Using computing resources closer to the end nodes (i.e., at the Edge or the Fog) can reduce overall end-to-end delay. Such an approach was followed in \cite{networkAwareScheduling}. Authors in this work designed a network-aware scheduling approach for container-based applications in Smart City deployment. Their approach builds on Kubernetes, enhancing the default scheduling mechanism available.

\subsection{Serverless and Functions-as-a-Service}
Serverless computing and \ac{faas} are emerging cloud architecture patterns that can provide significant cybersecurity benefits for \acp{cav} and \acp{cits}. The idea behind serverless is to ``focus on the application, not the infrastructure''. In a serverless model, applications run in short-lived, stateless containers triggered by specific events and fully managed by the cloud provider~\cite{serverless}. Resources are allocated dynamically and ``on-demand''. This contrasts traditional servers and virtual machines that run continuously regardless of utilisation.

For intelligent transportation systems, a serverless approach reduces the attack surface in several key ways:

\begin{itemize}
    \item Ephemeral function containers disappear after execution, minimising the ``window of opportunity'' for compromises (aligns with Req.~\ref{req:cscereq8}).
    \item Automated scaling removes resource management tasks vulnerable to misconfiguration (aligns with Req.~\ref{req:cscereq7})
    \item Stateless functions have no data at rest to exploit (aligns with Reqs.~\ref{req:cscereq8} and~\ref{req:ctefreq4}).
    \item Granular access controls can be applied per function (aligns with Req.~\ref{req:cscereq1}).
\end{itemize}

Additionally, \ac{faas} capabilities allow transportation services to scale elastically on demand. This is critical for maintaining \ac{qos} during unexpected traffic spikes or D\ac{dos} attacks. While an initial request may take longer to be handled than an application hosting platform, caching may enable subsequent requests to be handled within milliseconds.

By leveraging serverless and \ac{faas}, \ac{cits}, cybersecurity researchers and engineers can focus on developing discrete functions that can be scaled, updated, and secured independently. This flexible architecture aligns well with the dynamic nature of transportation systems. When combined with event-driven scaling, serverless allows \acp{cits} to be resilient and adaptive to evolving demands and threats.

\section{In-vivo Monitoring and Protection}\label{sec:inVivo}
The \ac{ctef} offers a comprehensive cybersecurity analysis for \ac{cits} components in isolated settings, but real-world \ac{cits} infrastructure protection is vital. \ac{ctef} is constructed to allow its tools to integrate seamlessly with real-world C-ITS systems. All the technologies described in Sec.~\ref{sec:technologies} can facilitate this integration and swiftly adapt new cybersecurity measures and ML solutions, ensuring minimal transition time for actual \ac{cits} systems. Furthermore, adopting maturity models ensures the selection and integration of high-quality solutions, reducing the risk of potential damage from faulty software or hardware. Incorporating ``Digital Twins''~\cite{digitalTwin1} further boosts the quality of solutions, ensuring the security, resilience, interoperability, and other fundamental requirements of a \ac{cits}. These tools also offer insights into the system's capabilities and help mitigate potential risks or adverse outcomes. The following sections describe how \ac{ctef} tools can be used in the real-world setup.

\subsection{Maturing Cybersecurity Tools}\label{subsec:maturingTools}
In conjunction with its operation as a testing and evaluation facility, \ac{ctef} can also provide its services (testing environments and \ac{secaas} platforms) for the development of cybersecurity incubators to perform research, development and maturity of new \ac{cits} cybersecurity tools and algorithms (aligns with Req.~\ref{req:cscereq4}). Starting with risk assessment frameworks (like ISO 27001~\cite{ISO27001}), the risks identified can be mitigated, leading to mature, real-world-ready solutions. 

The maturity readiness should be considered for the different tools and frameworks incubated. Well-known frameworks could be used for that like Capability Maturity Model Integration (CMMI)\footnote{Capability Maturity Model Integration (CMMI): \url{https://www.isaca.org/enterprise/performance-improvement-solutions}}, or the Cybersecurity Capability Maturity Model (C2M2)\footnote{Capability Maturity Model (C2M2) \url{https://www.energy.gov/ceser/cybersecurity-capability-maturity-model-c2m2}}. Internal frameworks could also be developed that better align with \ac{cits} organisations and systems. For example, authors in~\cite{maturingTools} describe a maturity model that can evaluate the performance of an organisation in a continuous and level-based way. The works~\cite{maturingModels1,maturingModels2} provide a good foundation for existing (or the lack of) end-to-end cybersecurity maturity assessment frameworks and ways forward to extend this area.

Using virtualised cloud-based cybersecurity platforms and simulated \ac{cits} environments allows easier deployment of cybersecurity testbeds where new tools and techniques can be developed and tested. The new tools will be matured within \ac{ctef}’s cybersecurity incubators to verify that they operate in accordance with the \ac{cits} operational requirements, as described in the in-situ testing and analysis of \ac{ctef}, before integrated into \ac{ctef}’s main \ac{secaas} platform.

\subsection{Cybersecurity of the Real-world Infrastructure}\label{subsec:realWorldInfra}
Both the in-vitro and the in-situ testing and evaluation tools and services that compose the \ac{ctef} will follow a \ac{saas} approach; thereby, they will provide the scalability and flexibility needed to be applied to larger systems such as a real-world \ac{cits}. Furthermore, since \ac{ctef} tools will operate with \ac{cits}’s operational requirements in mind (e.g., protocols, data flows, delays, etc.), they can be easily integrated into the real world. The only difference between \ac{ctef} and the real world, apart from the size of the system, is the need for continuous monitoring and security evaluation in a \ac{cits} environment due to the highly dynamic nature of the system and the ever-changing threat landscape (aligns with Reqs.~\ref{req:cscereq5} and~\ref{req:cscereq5}). 

For that purpose, localised \ac{siem} systems must be overlooking the real-world \ac{cits}, collaborating with localised \ac{ctef}-like implementations in a distributed manner. Each \ac{siem} will inform a localised \ac{soar} system responsible for triaging the data and responding to threats by taking remediation steps (aligns with Req.~\ref{req:cscereq5} and~\ref{req:cscereq7}). To allow organisation-wide strategy-driven decision-making, a strategic \ac{soar} will aggregate information from local \acp{soar} for monitoring purposes and also inform the organisation’s strategy back to the local \ac{soar}. An example of such an approach can be found in~\cite{siemSoar}, where a \ac{soar} dynamically deploys behavioural honeypots to enhance network intrusion detection and security measures. The above approaches can ensure that tools and strategies are implemented under a comprehensive risk management strategy (as discussed in~\cite{riskManagement}).

\begin{figure}[t]     
\centering
\includegraphics[width=1\columnwidth]{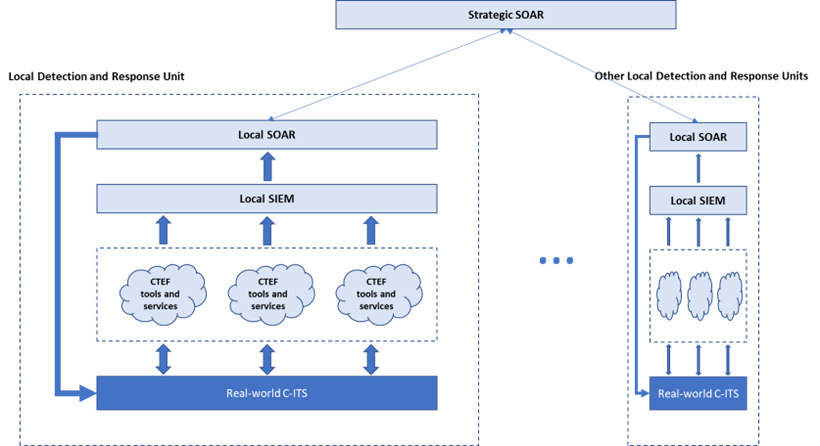}
    \caption{Cybersecurity of the real-world \ac{cits} infrastructure.}
    \label{fig:siemSoar}
\end{figure}

Fig.~\ref{fig:siemSoar} illustrates our view of the cyber protection of real-world \acp{cits}. In conjunction with the proposed \ac{ctef}, they will deliver holistic and continuous cybersecurity to \ac{cits} in a distributed, cloud-based manner. This approach will enable the use of the \ac{ctef} testing and evaluation tools outside \ac{ctef}, extending their exploitability.

\subsection{Digital Twins and Digital Network Oracles}\label{subsec:digitalTwins}
Sec.~\ref{subsec:insitu} discussed the importance of reconfigurable \ac{cits} test environments, the importance of the data, and the logging/measuring capabilities that should be incorporated in \ac{csce}. As it is evident, evaluating potential solutions in the real world could be very dangerous and raise privacy concerns. Therefore, the alternative solution is using ``Digital Twins''~\cite{digitalTwin1}. The term describes a virtual simulation of the physical world and assets -- whether it is a vehicle, the road infrastructure, or just a street of a major city (aligns with Req.~\ref{req:cscereq2}). 

Through data and feedback from both simulations and the real world, a Digital Twin can develop autonomy, learn from and reason from its environment. A Digital Twin is a \textit{``digital living simulation''} that brings all the data and models together, attempts to replicate salient features of the real world, and updates itself from multiple sources to represent its physical counterpart. There are already various attempts in the literature to capture different aspects of a \ac{cits}, such as~\cite{digitalTwin1} that describes a reference information sharing model,~\cite{digitalTwin2} that presents a lightweight traffic situation awareness Digital Twins, etc. 

Recently, the concept of ``Digital Network Oracles'', i.e., an augmented ``Digital Twin'', was introduced~\cite{driveSimulator}. The added capability refers to a particular architecture, which allows for the Digital Twin to be queried sequentially, to which it responds with the state of the ``virtual world''. With such capabilities, the Digital Twin can train iterative learning (e.g., Reinforcement learning) controllers for inferential and prediction tasks. 

A Digital Network Oracle is a virtual representation of physical assets in \acp{cpps} and can comprise several Digital Twins. If a Digital Twin is equipped with the following four characteristics: synchronisation with the physical assets, active data acquisition, the ability to simulate, and sequential bidirectional interaction with intelligent agents, the definition of this replica is transformed into a Digital Network Oracle. An oracle deployed in \ac{csce} should have the following features:

\begin{itemize}[left=0.5em]
    \item Automated detection of changes in scenarios in the real world and dealing with missing information in the digital domain.
    \item Automated detection of interdisciplinary dependencies and consistency checking of mechanics, electrics and software domain changes.
    \item Automated adaptation and changes in the models of the Digital Twin.
\end{itemize}

Moreover, an important feature required is the cross-domain (or cross-asset) synchronisation. Systems or sub-systems simulated in a Digital Twin do not adhere to the same simulation steps and are rather challenging to synchronise. Some existing synchronisation approaches have been demonstrated recently in different fields. For example, the cross-domain relationships of different assets were evaluated using semantic technologies~\cite{syncDemonstration} or using lifecycle management systems~\cite{syncDemonstration2} to edit the existing sub-models within a Digital Twin. To the best of our knowledge, these ways are partially or not automated and cannot be easily applied to the dynamic environments of a \ac{cits}. 

Lastly, regarding active data acquisition, such a system should require:

\begin{itemize}[left=0em]
    \item Acquisition of operational data from physical assets and processing and analysing the data.
    \item Provide various assistance functions, such as diagnosis and predictive quality.
    \item The ability to sample process data in a highly accurate manner and to assign it to the corresponding Digital Twin.
    \item Data unification or data curation.
\end{itemize}

Software modules and decentralised systems could be employed for such an approach. Furthermore, standardised connector interfaces ensure that shared data are being generated in the field, and standardised collectors (i.e., software modules) can collect the data in the Cloud or the Edge. Two aspects of a Digital Network Oracle, i.e., the ability to simulate and the bidirectional interaction with intelligent agents (e.g., in a Reinforcement Learning context), are also paramount. Examples of that can be found in~\cite{rlDigitalTwin,driveSimulator}. 

Digital Twins are expected to lead the experimentation on digital infrastructures in the near future. However, the complexity introduced when designing and developing such systems is tremendous. Especially when connecting different Digital Twins to create national Digital Twins able to emulate extensive complex scenarios, standard good practices should be followed between the different tools. 

\section{Challenges and Vulnerabilities}\label{sec:challenges}

The emerging technologies discussed in earlier sections undeniably enhance the capabilities of \acp{cits} and contribute to a more comprehensive cybersecurity approach. Yet, introducing nascent technologies often brings new challenges and potential vulnerabilities. These issues need addressing before full integration into any system. Transitioning from controlled environments, like the \ac{ctef}, to real-world applications presents its own set of challenges. This section delves into the challenges and solutions associated with the technologies above, standardisation efforts, and the vulnerabilities that might arise when applied in real-world scenarios.

\subsection{Wireless Communication Plane Challenges}
\begin{figure}[t]     
\centering
\includegraphics[width=1\columnwidth]{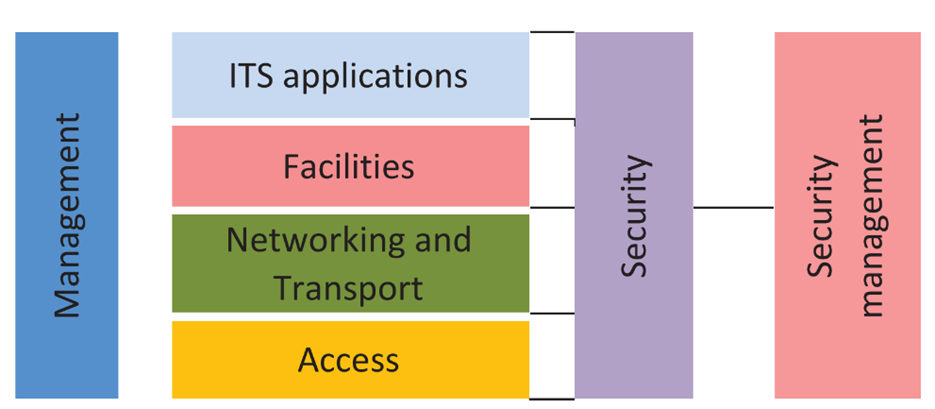}
    \caption{\ac{cits} security layers architecture~\cite{etsi10940}.}
    \label{fig:etsiStandard}
\end{figure}

Sec.~\ref{subsec:communicationDomains} touched upon the communication domains within a \ac{cits}, and Sec.~\ref{sub:citsSystemDesign} discussed the architecture of its future wireless plane. The wireless standards mentioned are grounded in the renowned OSI model. The ETSI TS 102 940 standard~\cite{etsi10940} delineates security requirements on a per-layer basis (Fig.~\ref{fig:etsiStandard}). This suggests the implementation of multiple security measures to safeguard a system. Additionally, the ETSI TS 102 942~\cite{etsi102942} standard offers a structure for access control in \ac{cits}, encompassing authentication, authorisation, and cryptographic certificate issuance.

Threat analysis is mandated for all \ac{v2x} communications. Such analyses have been proposed in past works such as~\cite{threatAnalysis}. Broadly, the attack surfaces correspond with the communication domains described in Sec.~\ref{subsec:communicationDomains}. Each domain has its unique characteristics and requirements, leading to unique vulnerabilities. Exploiting these vulnerabilities can lead to the attacks presented in Tab.~\ref{tab:attacks}. Existing literature presents standards that address the security of specific technologies and protocol stacks (e.g., ETSI  TS 133 501~\cite{etsi133501} and \ac{3gpp} 33.501~\cite{3gpp33501} deal with the \ac{3gpp} 5G system security), while others (e.g., ETSI TS 102 940~\cite{etsi10940},  ETSI TS 102 942~\cite{etsi102942}), deal more with the communication domains within a \ac{cits}. To encapsulate the challenges:

\begin{itemize}[left=0.5em]
    \item Varied authentication and authorisation levels are essential for \ac{cits} stations to access networks and services. 
    \item Point-to-point communications require confidentiality. Broadcast messages do not have specific requirements.
    \item Privacy measures, like pseudonym changes, for preventing tracking of \ac{cits} stations.
    \item Access control involving initialisation, enrolment and authorisation credentials ensures only trusted stations in the network.
    \item Trust and enrolment guidelines emphasise secure storage of keys and cryptographic operations.
    \item Security services, such as integrity and replay protection, are layered using defined access points.
    \item A \ac{pki} architecture must manage credentials, certificate trust lists, and a Misbehavior Authority for detecting and revoking misbehaving devices.
\end{itemize}

More details about the standards related to the wireless communication planes and the in-vehicle communication domains can be found in~\cite{standardWireless,standardsInVehicles}.

\subsection{Open-source Software Challenges}\label{sub:opensource}
While the prevailing opinion is that well-maintained open-source solutions are typically more secure, their open nature implies that vulnerabilities are also accessible by malicious actors. If not updated regularly, these vulnerabilities can be exploited. Authors in~\cite{ossRiskFactor} provide an overview of the software product quality problems, security issues and certain challenges confronting the \ac{oss}. The challenges associated \ac{oss} include:

\begin{itemize}[left=0.5em]
    \item \textbf{Public Exposure of Vulnerabilities}: Issues and vulnerabilities are often publicly listed on databases like the National Vulnerability Database\footnote{NIST: National Vulnerability Database https://nvd.nist.gov/}. Organisations that do not address these promptly can become targets of malicious actions.
    \item \textbf{Intellectual Property Risks}: Open-source components might inadvertently infringe on intellectual property rights due to the lack of standard commercial controls. 
    \item \textbf{Operational Inefficiencies}: Organisations might struggle to track and update open-source components, leading to potential risks.
    \item \textbf{Developer Malpractices}: Open-source components might become outdated or might not undergo rigorous testing. Poor coding habits can result in defective code that is hard to update or monitor.  
\end{itemize}

To mitigate these challenges, organisations should adopt to \acp{samm} as outlined in existing literature~\cite{softwareMaturity} or in publicly available open frameworks (e.g., OWASP \ac{samm}~\cite{softwareMaturity2}). These tools evaluate and improve software security practices, building balanced software security assurance and demonstrating security assurance improvements through an iterative process.

When deploying \ac{oss}, it is crucial to exercise extra caution. The unique nature of \ac{oss} demands specialised solutions to identify vulnerabilities, like \ac{sca} tools (or vulnerability scanners as otherwise known). Moreover, adhering to best practices related to software inventories, licensing, and continuous threat monitoring is vital throughout the \ac{sdlc} and software lifecycle~\cite{sca}. As discussed, software vulnerability detection can leverage traditional \ac{sca} tools or employ advanced technologies, including \ac{ml} algorithms, for both source~\cite{vulnerabilityDetectionSource} and binary~\cite{vulnerabilityDetectionBinary} code analysis. Just like with proprietary software, security responsibility should not rest solely with the vendor, allowing for transparent security patch resolutions.

\subsection{Security and Authorisation Challenges}
Security and authorisation in \ac{cits} primarily focus on ensuring the safe and reliable exchange of information among interconnected vehicles and infrastructure. This must be achieved while preserving the privacy and trustworthiness of the participating entities. As technology rapidly evolves and networked systems grow in scale, these challenges grow in complexity. We can categorise these challenges as follows:
\begin{itemize}[left=0em]
    \item \textbf{Regulation and Policy Adaptation}:  Crafting detailed security policies specifically for \acp{cits} and addressing any shortcomings or gaps in current standards. 
    \item \textbf{Identity Management and Privacy}: Handling the complex dynamics of digital certificates, from their issuance to revocation, and verifying the identity of devices and components within a \ac{cits}.
    \item \textbf{Resource Constraints and Scalability}: Ensuring processes, such as certificate revocation, remain efficient and prompt, even as the system expands.
    \item \textbf{Decentralisation and Dynamic Trust Evaluation}: Shifting towards decentralised infrastructure solutions and dynamically adjusting participant trust levels to maintain consistent system integrity.
\end{itemize}

The European Commission has tasked the \ac{cits} Platform with defining a security policy~\cite{europeCommisionCITS} to bolster the security of \ac{cits}. This policy governs the use of certificates, especially concerning short-range communications between vehicles, pedestrians, and the infrastructure network. ETSI standards already outlined a mechanism for disseminating and revoking certificates~\cite{etsi102942,etsi103097}. Existing standard limitations have been the subject of various research endeavours. For instance, the study in~\cite{openArchitecture} describes a novel architecture and a new trust model for a root \ac{ca}. In \ac{cits}, \ac{ca} is an entity that issues digital certificates and can be an integral part of the infrastructure network. Moreover, the \ac{ca} is responsible for revoking certificates associated with compromised or misbehaving entities. The study in~\cite{openArchitecture} combines certifications, web tokens, and transport layer security to guard against eavesdropping, malicious alterations, and \ac{mitm} attacks. Proposing a \ac{pki} supported by short-living tokens that are never reused minimises the impact of stolen tokens.

Similarly, vehicle identities can be substituted with multiple abstract short-lived identifiers, i.e., pseudonyms. The study in~\cite{pseudonymSchemes} explores this concept. The \ac{ca} issues these pseudonym credentials and verifies a vehicle's eligibility to share data by retaining its ID. This ensures location privacy since consecutive messages are signed using different pseudonyms. However, this approach demands more computational resources, potentially affecting system scalability and the efficiency of the revocation process. The Pseudonymous PKI (PPKI) method was examined in a smart grid context~\cite{certRevocation}, revealing that a single revocation scheme might not meet the overhead and security requirements of all smart grid applications.

Considering the decentralised infrastructure solutions proposed for future \acp{cits}, efficient decentralised \ac{pki} solutions are recommended. For example, authors in~\cite{decentralisedPKI} describe a learning game involving vehicles that utilise certificates distributed amongst themselves. These certificates can be revoked if a trust value drops below a set threshold. The environment rewards or penalises vehicles based on their actions in the game.

\subsection{Virtualisation \& Container Orchestration Challenges}
Virtualisation technologies, such as 5G \ac{oran}, enable mapping virtual machines (VMs) and containers to physical resources. This facilitates designing and deploying novel \ac{secaas} algorithms and features for future \ac{cits}~\cite{cloudNativeSolvesSecurity}. However, this new architecture poses security concerns, especially when users lose physical control over their computation and data. The primary attack vectors include:
\begin{itemize}[left=0.5em]
    \item \textbf{Architectural attacks}: 
    These exploit the abstraction layer between physical hardware and virtualised systems. For instance, VMs on the same network can be vulnerable to simultaneous attacks, simplifying unauthorised access. 
    \item \textbf{Hypervisor-based attacks}: These target hypervisor or container orchestration software vulnerabilities, leading to issues like resource exhaustion or VM sprawl.
    \item \textbf{(Mis)configurations}: 
    Cloning or copying images in a virtual environment can inadvertently deploy unwanted infrastructure components. This can introduce configuration drifts, making managing and securing the environment harder.
\end{itemize}

Mitigating these risks for VM hypervisors involves implementing policies for VM lifecycle management, controlling VM image creation and use, and ensuring quick recovery to a secure (initial/clean) state~\cite{hypervisors}. The security of the hypervisor is crucial, and managing rapidly deployed environments is vital. Similar principles apply to containers and their orchestrators.

For 5G-enabled computing solutions, such as \ac{oran}, the confidentiality of 5G services remains intact due to the distinction between RAN and core services. As described in~\cite{containerSecurity}, the data exchanged between containers undergo encryption and decryption at each function. This is also highlighted in security standards by \ac{3gpp}~\cite{5gSecurity,3gpp33501}. The introduction of 5G \ac{oran} overcomes previous hardware and vendor lock-in concerns, promoting a more open hardware future. The standardisation of \ac{oran} also addresses risks related to poorly managed open-source solutions (as introduced in Sec.~\ref{sub:opensource}).

However, while \ac{oran} and cloud-native implementations offer significant benefits (Sec.~\ref{subsec:cloudNative}), the added abstraction layers necessitate enhanced security. Ensuring secure interoperability in a complex multi-vendor environment is essential. Configuration errors can expose mission-critical resources and application traffic. Moreover, API management and security become paramount~\cite{apiSecurity}. Both the host systems and applications within containers must be safeguarded from threats like privilege escalation attacks~\cite{containerEscape}. Finally, applications inside the containers should also be protected from external risks (e.g., ransomware gaining access to a container)~\cite{containerSecurity}.

\subsection{Vulnerabilities in Serverless Architectures}
In serverless architectures, providers like AWS Lambda or Google Cloud Functions typically handle the security of all cloud components. However, this does not exempt developers from all responsibilities. Key implications of adopting serverless architectures include:
\begin{itemize}[left=0em]
    \item \textbf{Data Leakage}: Serverless functions are stateless, leading to sensitive data being stored externally and increasing leakage risks during data transfers.
    \item \textbf{Multi-tenancy}: Without dedicated servers for each service or user, functions run on shared resources. Misconfigurations can expose sensitive data or compromise system security.
    \item \textbf{Increased Attack Surface and Complexity}: Fragmenting applications into multiple serverless functions can expand potential attack vectors. Additionally, the ephemeral nature of these functions limits the time developers have to identify and address issues, complicating malicious event-data injection countermeasures.
    \item \textbf{Third-party Dependencies}: Relying on third-party software (open source, libraries, packages, etc.) in serverless development can be challenging to manage and increase the risk of inadequate security testing.
\end{itemize}

The inherent multi-tenancy of serverless computing introduces unique security threats. Features like pay-as-you-go and automatic scalability, while advantageous, can be exploited in \ac{dos} attacks, leading to unexpected costs for application owners~\cite{serverlessChallenges}. Many challenges in serverless computing mirror those in virtualised and containerised applications. The study in~\cite{serverlessChallenges} categorises serverless computing's security challenges into four domains: 1) resource isolation, 2) security monitoring, 3) security control, and 4) data protection. The authors also provide a detailed analysis of the risks in each domain and suggest potential mitigation strategies.

In conclusion, serverless architectures promise cost savings, scalability, and administrative ease, but they also introduce new security challenges. To safeguard serverless applications, continuous monitoring and automated security tools are paramount. Best practices include refining function permissions, enhancing logging and monitoring mechanisms, and ensuring robust data encryption.

\subsection{Challenges with the Real-world Integration}
The technologies discussed in Sec.~\ref{sec:technologies} aim to seamlessly integrate with real-world \ac{cits}. By introducing abstraction, virtualisation, and containerisation, we can separate cybersecurity applications and \ac{cits} services from hardware dependencies. This framework facilitates the scaling of envisioned solutions, ensuring that real-world \ac{cits} can swiftly adopt new cybersecurity frameworks and \ac{ml}-based solutions. Such an approach guarantees production-grade operation within the \ac{csce}.

Furthermore, the maturity model introduced in Secs.~\ref{subsec:maturingTools} and~\ref{subsec:realWorldInfra} ensures that solutions, when deployed in the real world, are of high quality, thus minimising potential catastrophic impacts from flawed software or hardware. In essence, a maturity model gauges the progress in embedding security into daily operations and the strategic undertakings of \ac{cits}. The Digital Twins and Digital Network Oracles introduced in Sec.~\ref{subsec:digitalTwins} further bolsters the solutions the \ac{csce} provides. They help establish key requirements of a \ac{cits}, such as security, resilience, and interoperability. 

However, despite the advantages of these technologies and strategies, several challenges arise in real-world integration:

\subsubsection{Privacy}
\ac{cits} applications often process user identification data (e.g., usernames, registration plates, etc.), which must be protected against unauthorised access and comply with local privacy regulations (such as \ac{gdpr}). Even applications that do not process personal information can inadvertently de-anonymise users when data is correlated with other datasets. The challenge lies in ensuring that cybersecurity tools designed for or within \ac{ctef}, consider data privacy when applied to real-world scenarios. Authors in~\cite{humanDataInteraction} identify this as an issue in the EU Regulation on the Free Flow of Non-personal Data~\cite{euRegulation}. Data anonymity is a major issue that has to be investigated in the context of \ac{cits}. Within the proposed \ac{csce}, the challenge lies in ensuring that cybersecurity tools designed for or within \ac{ctef}, an isolated \ac{cits} environment, consider data privacy when applied to real-world scenarios.

\subsubsection{Operation}
\ac{cits} environments, with their diverse hardware, software, and network platforms, have strict operational requirements. As noted in~\cite{Fysarakis2017SECURITYCI}, many \ac{cits} services (e.g., a crash avoidance system) rely on communications with very low latency and other \ac{qos} characteristics. Traditional cybersecurity approaches may not be suitable due to the unique nature of \ac{cits}. This highlights the need to develop new cybersecurity tools tailored to \ac{cits}. \ac{ctef} can play a pivotal role in the research and development efforts required, providing space for incubating novel activities and testing existing ones for their applicability in a \ac{cits}.

\subsubsection{Resilience of a Real-world System}
The fusion of legacy and modern technologies in industrial control and enterprise IT, combined with emerging technologies within \ac{cav} platforms, leads to a new era of resilience challenges. The \ac{csce} plays a pivotal role in navigating these challenges, leveraging metrics like the Mean Time Between Failure and standards such as Safety Integrity Levels (SIL)~\cite{sil} and IEC 62061~\cite{iec62061}.

The evolving landscape of cyber threats demands a dynamic approach to resilience, especially in cybersecurity. Established standards, from STIX~\cite{stix}, provide a foundation for understanding threats, sharing intelligence and countering them. Cybersecurity models such as MITRE’s ATT\&CK framework~\cite{mitre} can help evaluate resilience concerning Threat Actors / Actor groups and their use of exploits/intrusion sets. The even-evolving data and the \ac{cav}'s platform robustness against attacks can be described in terms of Tools, Techniques and Procedures (TTPs). Adopting new technologies like ML introduces complexities, necessitating a more granular, component-level assessment of resilience.

In this interconnected age, a holistic approach to resilience is paramount. This encompasses the: 1) integration of old but stable and new but unpredictable technical solutions, 2) continuous evaluation of the resilience as the systems and threats evolve, 3) proactive strategies, aiming to defend and anticipate, adapt, and evolve in the face of emerging challenges, 4) stakeholder collaboration including technology providers, regulatory bodies, and end-users, and 5) training and awareness ensuring that those involved in the design, deployment, and operation of \ac{cits} platforms are well-trained and aware of the latest threats and best practices. 

Building resilience in real-world systems, especially as complex as \acp{cits}, is challenging but essential. It requires combining technical solutions, continuous evaluation, stakeholder collaboration, and human-centric approaches. As the world becomes more connected and technology continues to evolve, the importance of resilience will only grow.

\section{Conclusion}\label{sec:conclusion}
The ongoing digital transformation of transportation systems unveils a plethora of opportunities and challenges. At the centre of these challenges lies cybersecurity, which has become not only a technical priority but also a societal obligation. The \ac{csce} and \ac{ctef} frameworks outlined in this paper offer a structured approach to address these cybersecurity challenges, always with an eye on the seamless integration of emerging technologies into real-world scenarios. Our work underscored the importance of robust testing mechanisms and facilities, detailing the key technologies for providing the required functionality. We portrayed our envisaged requirements across the in-situ, in-vitro, and in-vivo domains of a prospective \ac{cits}. Additionally, we delved into the inherent challenges within each domain, offering insights on the adopted technologies or suggestions for future research pathways. A multi-layered security strategy is essential to counteract the diverse array of attack vectors. Integrating ML-powered intelligent agents can strengthen detection capabilities, offering a proactive attitude towards threat identification and mitigation. As the transportation sector gravitates towards real-world \ac{cits} deployments, continuous monitoring becomes paramount. Safeguarding \ac{cav} services is vital and guarantees their peak performance and continuous operation. Moving forward, the focus should remain on enhancing cybersecurity, fine-tuning detection methodologies, and broadening the spectrum of real-world testing scenarios.

\ack This work was supported in part by Toshiba Europe Ltd. and in part by the CAVShield project (grant no. 133898, UK Research and Innovation, Innovate UK).

\bibliographystyle{icstnum.bst}
\bibliography{bib}

\end{document}